\documentclass[a4paper,11pt]{article}

\usepackage[height=23.6truecm,width=16truecm,top=3.05truecm]{geometry}

\usepackage[skip=6pt,indent]{parskip}

\usepackage[utf8]{inputenc}
\usepackage{graphicx}
\usepackage{amsmath}
\usepackage{amssymb}
\usepackage{hyperref}
\hypersetup{colorlinks,citecolor=TokiwaIro,linkcolor=RuriIro,urlcolor=RuriIro,linktocpage}
\usepackage{cite}
\usepackage{braket}
\usepackage{bm}
\usepackage{bbm}
\usepackage{color}
\usepackage[svgnames,table]{xcolor}
\usepackage{float}
\usepackage{hhline}
\usepackage{multirow}
\usepackage{indentfirst}
\usepackage{latexsym}
\usepackage{mathtools}
\usepackage{cancel}
\usepackage{slashed}
\usepackage{colortbl}
\usepackage[mathscr]{euscript}
\usepackage{enumitem}
\usepackage{subcaption}
\usepackage{comment}
\usepackage{amsthm}

\usepackage{tikz-feynhand}

\numberwithin{equation}{section}

\allowdisplaybreaks

\definecolor{RuriIro}{rgb}{0.,0.28,0.60}
\definecolor{TokiwaIro}{rgb}{0.,0.39,0.16}
\definecolor{AkaneIro}{rgb}{0.72,0.16,0.18}

\definecolor{kblue}{rgb}{0,0.48,0.73}
\definecolor{kred}{rgb}{0.73,0.25,0}
\definecolor{kgreen}{rgb}{0.48,0.73,0}
\definecolor{dgreen}{rgb}{0.2,0.51,0.19}

\theoremstyle{plain}

\theoremstyle{remark}

\newcommand{\nn}{\nonumber}

\newcommand{\mc}{\mathcal}
\newcommand{\mr}{\mathrm}

\newcommand{\del}{\partial}

\newcommand{\tr}{\mathop{\mathrm{tr}}\nolimits}
\newcommand{\Tr}{\mathop{\mathrm{Tr}}\nolimits}

\renewcommand{\Re}{\mathop{\mathrm{Re}}}
\renewcommand{\Im}{\mathop{\mathrm{Im}}}

\newcommand{\EH}{\mathrm{EH}}
\newcommand{\GHY}{\mathrm{GHY}}
\newcommand{\Maxwell}{\mathrm{Maxwell}}

\newcommand{\U}{\mathrm{U}}

\newcommand{\SO}{\mathrm{SO}}
\renewcommand{\U}{\mathrm{U}}

\definecolor{dred}{rgb}{0.7,0.15,0.09}
\definecolor{dblue}{rgb}{0,0.12,0.64}
\definecolor{dgreen}{rgb}{0.2,0.51,0.19}

\definecolor{murasaki}{rgb}{0.42,0.19,0.61}

\begin{document}

\begin{titlepage}

\begin{flushright}
\end{flushright}

\vspace{1cm}

\begin{center}

{\LARGE \bfseries
Graviphoton mixing with KK modes in the 2D effective theory of charged black holes
}

\vspace{1cm}

\renewcommand{\thefootnote}{\fnsymbol{footnote}}
{%
\hypersetup{linkcolor=black}
Yoshihiko Abe$^{1,2,3}$\footnote[1]{yabe3@keio.jp},
\ 
Tetsutaro Higaki$^{4}$\footnote[2]{thigaki@rk.phys.keio.ac.jp},
\ and
Yu Miyauchi$^{4}$\footnote[3]{mu\_19990710@keio.jp}
}%
\vspace{8mm}

{\itshape%
$^1${Graduate School of Science and Technology, Keio University, Yokohama, Japan}\\
$^2${Keio University Sustainable Quantum Artificial Intelligence Center (KSQAIC), Keio University, Tokyo 108-8345, Japan}\\
$^3${Quantum Computing Center, Keio University, 3-14-1 Hiyoshi, Kohoku-ku, Yokohama, Kanagawa, 223-8522, Japan}\\
$^4${Department of Physics, Keio University, Yokohama 223-8533, Japan}
}%

\vspace{8mm}

\end{center}

\abstract{
We study the effects of massive Kaluza--Klein modes on the two-dimensional low-energy
effective theory obtained by the $S^2$ reduction of four-dimensional Einstein--Maxwell theory
on a Reissner--Nordstr\"om black hole background.
We retain the first excited ($l=1$) Maxwell Kaluza--Klein modes in addition to the massless sector.
The background electric flux mixes the Maxwell Kaluza--Klein scalars with the graviphotons, and tree-level exchange of the massive scalars generates a correction to the graviphoton kinetic term proportional to the square of the black hole charge.
Within the leading local approximation, we also perturbatively evaluate the one-loop correction to the graviphoton kinetic term and find no dependence on the black hole charge. We find that near extremality, the charge-dependent tree-level correction becomes
parametrically comparable to the classical graviphoton kinetic term.
}

\end{titlepage}

\renewcommand{\thefootnote}{\arabic{footnote}}
\setcounter{footnote}{0}
\setcounter{page}{1}

\tableofcontents

\section{Introduction}
\label{sec:Introduction}

Near-extremal charged black holes provide a tractable setting for studying quantum gravitational effects.
At extremality, the near-horizon geometry of a four-dimensional Reissner--Nordstr\"om (RN) black hole in Einstein--Maxwell theory is $AdS_2 \times S^2$, and spherical reduction yields a two-dimensional dilaton gravity theory coupled to U(1) and SO(3) gauge fields \cite{Iliesiu:2020qvm}.
Since the partition functions of two-dimensional gauge theories are exactly computable~\cite{Migdal:1975zf,Migdal:1983qrz,Blau:1991mp,Fine:1991ux,Witten:1991we,Witten:1992xu,Cordes:1994fc}, one can integrate out the U(1) and SO(3) gauge fields to obtain effective dilaton gravity.
In the near-horizon limit of a near-extremal black hole, this theory reduces to Jackiw–Teitelboim (JT) gravity~\cite{Jackiw:1984je,Teitelboim:1983ux,Almheiri:2014cka}.
The exact solvability of JT gravity has played a central role in recent progress on near-extremal black holes \cite{Maldacena:2016upp,Saad:2019lba,Iliesiu:2019xuh}. 
For example, Ref.~\cite{Iliesiu:2020qvm} shows how the apparent mass-gap problem for RN black holes~\cite{Preskill:1991tb,Maldacena:1998uz,Page:2000dk} is resolved.
These developments motivate a detailed study of the dimensionally reduced theory and its quantum corrections.

In dimensional reductions, massive Kaluza--Klein (KK) modes are often truncated when the characteristic energy scale, such as the black hole temperature, lies well below the KK scale, suppressing their thermal excitation. Within the approximations considered, 
previous studies found that massive KK modes modify neither
the leading low-temperature behavior of the near-extremal partition function \cite{Iliesiu:2020qvm} nor the qualitative semiclassical phase structure of RN--AdS black holes \cite{Abe:2026sai}.
Thermal suppression, however, does not imply that virtual massive KK modes can be neglected in the low-energy effective theory, 
because virtual exchange and quantum loops can modify the 
low-energy couplings.

Virtual effects of massive KK modes are particularly relevant for sphere reductions in which the SO(3) graviphotons are retained.
In general, retaining only a finite set of KK modes does not define a consistent truncation because the retained fields can source higher KK modes that have been omitted.\footnote{Consistent sphere reductions to a finite set of lower-dimensional fields are nevertheless known in particular theories, where the symmetry structure and field content allow the truncated sector to close consistently~\cite{deWit:1986oxb,Cvetic:2000dm,Gibbons:2003gp,Ciceri:2023bul}.}
More generally, heavy modes can affect the interactions and observables of the light sector even when the heavy mode masses are parametrically above the energy scale of interest~\cite{Randjbar-Daemi:2006ada,Akhoury:2007bb}.
The possible effects of heavy modes motivate a direct study of
how massive KK modes modify the massless sector couplings in a charged black hole background.

In this paper, we study the $S^2$ reduction of four-dimensional 
Einstein--Maxwell theory on an asymptotically flat RN black hole background.
In addition to the massless sector, we retain the $l=1$ massive Maxwell KK modes and evaluate the resulting corrections to the gauge kinetic terms. The background electric field mixes the $l=1$ Maxwell KK scalars with the SO(3) graviphotons, and tree-level scalar exchange generates a correction to the graviphoton kinetic term proportional to the square of the black hole charge.
The corresponding one-loop correction to the graviphoton kinetic term is independent of the black hole charge within the leading local approximation.
Near extremality, the charge-dependent tree-level correction becomes parametrically comparable to the classical graviphoton kinetic term.
Thus, the massive KK sector gives a non-negligible correction to the SO(3) gauge coupling when graviphoton fluctuations are included.

This paper is organized as follows. 
Sec.~\ref{sec:4D-charged-black-holes-and-2D-reduced-theory} reviews the asymptotically flat RN black hole solution and the corresponding two-dimensional reduction.
In Sec.~\ref{sec:Effective-action}, we integrate out the \(l=1\) massive Maxwell KK modes, derive the resulting corrections to the effective action, and assess when they are suppressed relative to the classical terms.
We discuss the interpretation and limitations of our results in Sec.~\ref{Sec:Conclusion}.

\section{4D charged black holes and 2D reduced theory}
\label{sec:4D-charged-black-holes-and-2D-reduced-theory}

In this section, we briefly review four-dimensional charged black holes in Einstein–-Maxwell theory, starting from 
the Euclidean action and the RN black hole solution. 
Following~\cite{Iliesiu:2020qvm}, we then present the two-dimensional reduced theory for the massless metric and electromagnetic fluctuations about the RN black hole background. 
The reduced theory consists of two-dimensional dilaton gravity, $\SO(3)$ Yang--Mills theory, and Maxwell theory.

\subsection{4D spherically symmetric charged black hole}

We begin with four-dimensional Einstein–Maxwell theory on the RN black hole background. The Euclidean action\footnote{
	We adopt units where $c = \hbar = k_B = 1$.
    The Lorentzian time $t_L$ and the Euclidean time $t_E$ are related by the Wick rotation $t_L=-it_E$. 
    In our notation, the Lorentzian and Euclidean gauge fields, $A_\mu^L$ and $A_\mu^E$, are related by $A_{t_L}^L=iA_{t_E}^E$ and $A_m^L=A_m^E$, where $m$ denotes a spatial index.
    Throughout this paper, we omit the label $E$ for simplicity.
} is given by 
\begin{align}
	I[g_{\mu\nu}, A_\mu] = & I_\EH[g_{\mu\nu}] + I_\GHY[g_{\mu\nu}] + I_\Maxwell[A_\mu,g_{\mu\nu}],
	\label{eq:Einstein-Maxwell-action}
	\\
	I_\EH[g_{\mu\nu}] &= - \frac{1}{16 \pi G_N} \int_{\mc{M}} d^4x \sqrt{\det g_{(4)}} R_{(4)},
	\label{eq:Einstein-Hilbert-action}
	\\
	I_\GHY[g_{\mu\nu}] &=- \frac{1}{8 \pi G_N} \int_{\del \mc{M}} d^3x \sqrt{\det \gamma_{(3)}} K_{(3)},
	\label{eq:Gibbons-Hawking-York-term}
	\\
	I_\Maxwell[A_\mu, g_{\mu\nu}] &= \frac{1}{4e^2} \int_{\mc{M}} d^4x \sqrt{\det g_{(4)}} F_{\mu\nu} F^{\mu\nu}.
	\label{eq:Maxwell-action}
\end{align}
Here, $\mc{M}$ denotes the four-dimensional Euclidean spacetime, and
$\del \mc{M}$ denotes a boundary at fixed $r$ outside the event horizon.
$G_N$ denotes the four-dimensional Newton constant.
Greek indices $\mu,\nu$ run over the four coordinates of the Euclidean spacetime $\mc{M}$.
The induced metric on $\del \mc{M}$ is given by $\gamma_{\mu\nu} = g_{\mu\nu} - n_{(4)\mu} n_{(4)\nu}$, where $n_{(4)\mu}$ is the unit normal vector of the boundary.
$R_{(4)}$ is the scalar curvature of $\mc{M}$.
The extrinsic curvature of $\partial\mathcal{M}$ is
$K_{(3)}:=\nabla_\mu n^\mu_{(4)}$, where $\nabla_\mu$ denotes the
Levi-Civita covariant derivative associated with the four-dimensional
metric $g_{\mu\nu}$.
Here, $A_\mu$ is the U(1) gauge field, $F_{\mu\nu}=\partial_\mu A_\nu - \partial_\nu A_\mu$ its field strength, and
$e$ the four-dimensional gauge coupling.
For details, see also Ref.~\cite{Abe:2026sai}.

The spherically symmetric electrically charged black hole is characterized by its mass $M$ and electric charge $Q$.
The metric and gauge field configurations are given by 
\begin{align}
	g_{\mu\nu}^{\text{RN}}dx^\mu dx^\nu &= f(r) dt^2 + \frac{dr^2}{f(r)} + r^2 h_{\alpha\beta} dy^\alpha dy^\beta,
    \quad
    f(r):= 1 - \frac{2 G_NM}{r} + \frac{e^2}{4\pi}\frac{G_NQ^2}{r^2},
	\label{eq:4-dim-Riesnner-Nordstrom}
	\\
	A^{\text{RN}}_{\mu} dx^\mu &= i\frac{e^2}{4 \pi}\frac{Q}{r} dt,
	\label{eq:4-dim-gauge-field}
\end{align}
where $y^\alpha$ ($\alpha=3,4$) are coordinates on $S^2$, with the metric $h_{\alpha\beta} dy^\alpha dy^\beta :=  d \theta^2 + \sin^2 \theta d\phi^2$.
The factor $i$ in Eq.~\eqref{eq:4-dim-gauge-field} ensures that the Euclidean solution matches the Lorentzian solution after Wick rotation.
We use the configuration in Eqs.~\eqref{eq:4-dim-Riesnner-Nordstrom} and \eqref{eq:4-dim-gauge-field} as the RN black hole background in the following analysis.
The horizon radii are the roots of $f(r)=0$:
\begin{align}
    r_\pm :=  G_NM \pm \sqrt{ G_N^2M^2- \frac{e^2}{4\pi}G_N Q^2},
    \label{eq:r_H&r_pm}
\end{align}
where $r_+$ and $r_-$ are the outer and inner horizon radii, respectively.
Let $r_H\equiv r_+$ denote the outer horizon radius.
The black hole temperature is determined by requiring regularity of the Euclidean geometry at the horizon, which fixes the periodicity of the Euclidean time coordinate as
\begin{align}
    t \sim t+\beta,
    \qquad
    \beta = 4\pi \left( \frac{df}{dr}(r_H)\right)^{-1}.
    \label{eq:BH-inverse-temperature-4D}
\end{align}
The inverse period $\beta^{-1}$ gives the black hole temperature,
\begin{align}
    T
    = \frac{1}{4\pi} \frac{df}{dr}(r_H)
    = \frac{r_H-r_-}{4\pi r_H^2}.
    \label{eq:BH-temperature-4D}
\end{align}
Black holes satisfying $0<r_H-r_-\ll r_H$ have low but nonzero temperatures and are called near-extremal black holes.
For a near-extremal black hole, Eqs.~\eqref{eq:r_H&r_pm} and \eqref{eq:BH-temperature-4D} imply
\begin{align}
    r_H^2 \sim G_N^2M^2 \sim \frac{e^2}{4\pi}G_NQ^2.
    \label{eq:near-extremal-condition}
\end{align}

\subsection{2D reduced theory with massive modes truncated}

To study quantum effects, 
we consider metric and electromagnetic fluctuations, $\delta g_{\mu\nu}$ and $\delta A_\mu$, about the RN black hole background $g_{\mu\nu}^{\text{RN}}$ and $A_\mu^{\text{RN}}$.
Using spherical symmetry, we decompose the fluctuations into two-dimensional KK modes.
We retain the massless KK modes, truncate the massive modes, and 
integrate over $S^2$ to obtain the two-dimensional massless sector action.

We take the four-dimensional spacetime $\mc{M}$ to factorize into a two-dimensional spacetime $\Sigma$ and an internal sphere $S^2$:
\begin{align}
    \mc{M} = \Sigma \times S^2,
    \qquad 
    \del \mc{M} = \del \Sigma \times S^2.
    \label{eq:geometric-assumption}
\end{align}
We denote coordinates on $\Sigma$ by $x^a$ ($a=1,2$).

The four-dimensional fluctuations $\delta g_{ab}$ and $\delta A_a$ generally depend on both $x$ and $y$.
In the standard $S^2$ reduction, the fluctuations are expanded in real-valued spherical harmonics $Y^{\mathbb{R}}_{(l,m)}(y)$, which are eigenfunctions of the $S^2$ Laplacian labeled by the angular momentum quantum numbers $l$ and $m$.
The spherical harmonics encode the $y$ dependence, while the expansion coefficients depend only on $x$ and become two-dimensional KK fields on $\Sigma$. KK modes with nonzero angular momentum generally acquire $l$-dependent masses. 
We focus on the low-energy regime in which the relevant scale, such as the black hole temperature, is well below the KK mass scale;
in this section, we retain only the massless modes.

The electromagnetic field is parametrized as
\begin{align}
    A^\text{RN}_\mu dx^\mu + \delta A_\mu dx^\mu 
    &= \frac{1}{\sqrt{4\pi}} a_a(x) dx^a,
    \label{eq:delta-a-massless}
\end{align}
where, for notational simplicity, we denote by $a_a$ the sum of the RN black hole background $\sqrt{4\pi}A_a^{\text{RN}}(x)$ and its $l=0$ fluctuation $\delta a_a(x)$.
The factor $1/\sqrt{4\pi}$ is introduced for convenience in the $S^2$ reduction.
$\delta a_a$ corresponds to the two-dimensional U(1) gauge field.\footnote{
    The two-dimensional U(1) gauge transformation is given by $a_a \mapsto a_a + \partial_a \theta$, where $\theta(x)$ is an infinitesimal gauge transformation parameter. This U(1) gauge symmetry originates from the four-dimensional transformation $A_\mu \mapsto A_\mu + \partial_\mu \Theta$ by setting $\Theta=\theta/\sqrt{4\pi}$, where $\Theta(x,y)$ is an infinitesimal gauge transformation parameter.
}
We refer to $\delta a_a(x)$ as the two-dimensional U(1) photon.
In the gravitational sector, the massless field includes not only $l=0$ modes but also $l=1$ ones:
\begin{align}
ds^2 =    g_{\mu\nu}^{\text{RN}} dx^\mu dx^\nu+\delta g_{\mu\nu} dx^\mu dx^\nu &= g_{ab} dx^a dx^b+ \Phi(x) h_{\alpha\beta}(y) e^\alpha e^\beta,
    \label{eq:delta-g-massless}
\end{align}
where the two-dimensional metric $g_{ab}$ contains both the two-dimensional part of the RN black hole background $g_{ab}^{\text{RN}}$ and its $l=0$ metric fluctuation along $\Sigma$.
The dilaton $\Phi(x)$ describes the size of $S^2$ and its $l=0$ fluctuation.
$e^\alpha$ is defined by
\begin{align}
    e^\alpha &:= dy^\alpha + B^I_a(x) \zeta_I^\alpha (y) dx^a,
    \label{eq:def-of-e}
\end{align}
where the graviphotons $B_a^I$ are the $l=1$ metric fluctuations along $S^2$, and $\zeta_I := \zeta^\alpha_I\partial_\alpha$ ($I=\pm1,0$) are the Killing vectors on $S^2$:
\begin{align}
    \zeta_{+1} =-\sin \phi \partial_\theta - \cot \theta \cos \phi \partial_\phi,
    \quad
    \zeta_{-1} = \cos \phi \partial_\theta - \cot \theta \sin \phi \partial_\phi,
    \quad
    \zeta_{0} = \partial_\phi.
    \label{eq:def-of-Killing}
\end{align}
The Killing vectors $\zeta_I$ generate the $\SO(3)$ isometry group and satisfy
\begin{align}
    [\zeta_I,\zeta_J]= -\sum_{K=\pm1,0}\epsilon_{IJK} \zeta_K.
\end{align}
The antisymmetric symbol is normalized by $\epsilon_{+1,-1,0}=+1$.
See App.~\ref{app:Einstein-Hilbert-action} for details.
To see how the SO(3) gauge symmetry keeps the graviphoton massless, consider the transformation
\begin{align}
    y^\alpha \mapsto y^\alpha + \lambda^I (x) \zeta_I^\alpha,
    \label{eq:SO(3)-transformation-y}
\end{align}
where $\lambda^I(x)$ is an infinitesimal transformation parameter depending on the two-dimensional coordinate.
In order for the one-form $e^\alpha$ defined by Eq.~\eqref{eq:def-of-e} to transform covariantly under the SO(3) isometry~\eqref{eq:SO(3)-transformation-y}, the graviphotons transform as
\begin{align}
    B^I_a \mapsto B^I_a + D_a \lambda^I,
    \quad
    D_a \lambda^I:= \partial_a \lambda^I 
    +\epsilon_{JKI} B_a^J \lambda^K.
    \label{eq:SO(3)-transformation-B}
\end{align}
Thus, the graviphotons are the SO(3) gauge fields, and $D_a$ is the corresponding covariant derivative.

Substituting Eqs.~\eqref{eq:delta-a-massless} and \eqref{eq:delta-g-massless} into the four-dimensional action~\eqref{eq:Einstein-Maxwell-action} and integrating over $S^2$ yields the massless sector action, consisting of two-dimensional dilaton gravity, 
SO(3) Yang--Mills theory, and U(1) Maxwell theory~\cite{Iliesiu:2020qvm}:
\begin{align}
    I_{\text{massless}}
    &= - \frac{1}{4 G_N} \int_{\Sigma} d^2x \sqrt{\det g} \biggl[
        \Phi \Bigl(
            R_{(2)} + \frac{2}{r_0^2}
        \Bigr) + \frac{3}{2 \Phi} (\partial\Phi)^2
    \biggr]
    - \frac{1}{2 G_N} \int_{\del \Sigma} dx \sqrt{\det \gamma} ~\Phi K_{(1)}
    \notag\\
    &\qquad 
    + \frac{r_0^2}{24G_N}\int_{\Sigma} d^2x \sqrt{\det g}~\Phi H^I_{ab}H^{Iab}
    + \frac{r_0^2}{4e^2} \int_{\Sigma} d^2x \sqrt{\det g} ~f_{ab} f^{ab},
    \label{eq:I_massless}
\end{align}
where $r_0$ is the compactification radius, and we have performed the Weyl transformation
\begin{align}
    g_{ab} \mapsto \frac{\Phi}{r_0^2}g_{ab}.
    \label{eq:Weyl-transformation}
\end{align}
The induced metric on $\partial\Sigma$ is $\gamma_{ab}:=g_{ab} - n_{(2)a}n_{(2)b}$, where $n^a_{(2)}$ is the unit normal.
$H_{ab}^I:= \partial_a B_b^I - \partial_b B_a^I + \epsilon_{JK}{^I} B_a^JB_b^K$ and $f_{ab}:=\partial_a a_b - \partial_b a_a$ are the SO(3) and U(1) field strengths, respectively.
$R_{(2)}$ is the scalar curvature for the two-dimensional metric $g_{ab}$, and $K_{(1)}$ is the extrinsic curvature for the induced metric $\gamma_{ab}$.
See App.~\ref{App:Details-of-dimensional-reduction} for details.
The two-dimensional classical solution in the massless sector with $H^I_{ab}=0$ reproduces the four-dimensional classical solution~\eqref{eq:4-dim-Riesnner-Nordstrom} and \eqref{eq:4-dim-gauge-field}~\cite{Abe:2026sai}.

Since the graviphotons carry angular momentum $l=1$, they can couple KK modes whose angular momenta differ by one.
In the charged black hole background, this coupling leads, in particular, to mixing between the massless graviphotons and the $l=1$ massive Maxwell KK modes, inducing corrections to the massless sector effective action.
At linear order, this mixing corresponds to the $l=1$
odd-parity coupling between gravitational and electromagnetic perturbations analyzed by Moncrief~\cite{Moncrief:1975sb}.
In the next section, we show explicitly how this mode mixing arises and clarify the conditions under which the contributions from the massive modes induced by this mixing can be suppressed.

\section{2D effective action for the massless modes}
\label{sec:Effective-action}

In this section, we include massive KK modes in the dimensional reduction and estimate their corrections to the massless sector effective action. Using the metric ansatz~\eqref{eq:delta-g-massless}, we omit massive gravitational KK modes and evaluate the tree-level and one-loop contributions from the Maxwell KK modes to the two-dimensional U(1) and SO(3) gauge couplings. 
In Sec.~\ref{sec:Corrections-from-the-massive-modes-to-the-massless-modes}, we restrict the explicit calculation to the \(l=1\) modes and retain only local bulk terms quadratic in the massless fluctuations, omitting higher-order, nonlocal and finite-temperature effects, boundary terms, and infrared running. Contributions from \(l\ge2\) modes are discussed separately in Sec.~\ref{sec:Corrections-from-the-l>1-massive-modes}.

\subsection{Kaluza--Klein reduction of the Maxwell theory on $S^2$}
\label{sec:Kaluza-Klein-reduction}

We consider the electromagnetic fluctuations with arbitrary dependence on the internal coordinates $y$.
Because the background is spherically symmetric, 
the electromagnetic fluctuations can
be expanded in spherical harmonics \cite{Michelson:1999kn,Donnelly:2015hxa,Iliesiu:2020qvm,Abe:2026sai} as 
\begin{align}
    \delta A_\mu (x,y) dx^\mu &=  \delta A_a(x,y)dx^a +  \delta A_\alpha (x,y) e^\alpha,
    \label{eq:A_a&A_alpha}
    \\
    \delta A_a(x,y) &= \delta a_a(x) Y_{(0,0)}^{\mathbb{R}} +  \sum_{(l,m)} a_{a}^{(l,m)}(x) Y^{\mathbb{R}}_{(l,m)}(y),
    \label{eq:4D-gauge-field-A_a-decomposition}
    \\
    \delta A_\alpha(x,y) &=  \sum_{(l,m)}\frac{1}{m_l} \Bigl[
    \tilde{\varphi}^{(l,m)}(x) \partial_\alpha Y^{\mathbb{R}}_{(l,m)}(y) + \varphi^{(l,m)}(x) \epsilon_{\alpha\beta}(y) \partial^\beta Y^{\mathbb{R}}_{(l,m)}(y)
    \Bigr],
    \label{eq:4D-gauge-field-A_k-decomposition}
\end{align}
where $\epsilon_{\alpha\beta}$ is the antisymmetric tensor on $S^2$.\footnote{
In $D$-dimensional Euclidean space, we take the antisymmetric tensor to satisfy
\begin{align}
    \epsilon_{12\cdots D} = +\sqrt{\det g_{(D)}}.
\end{align}
}
The real-valued spherical harmonics $Y^{\mathbb{R}}_{(l,m)}$ are eigenfunctions of the Laplacian on $S^2$:
\begin{align}
    -h^{\alpha\beta} \nabla_\alpha \nabla_\beta Y^{\mathbb{R}}_{(l,m)}=l(l+1)Y^{\mathbb{R}}_{(l,m)}.
    \label{eq:eigenvalue-eq-SH}
\end{align}
They are normalized as
\begin{align}
    \int_{S^2} d^2y \sqrt{\det h}~Y_{(l,m)}^{\mathbb{R}}(y) Y_{(l',m')}^{\mathbb{R}}(y)= \delta_{ll'} \delta_{mm'},
    \label{eq:orthogonal-condition-SH}
\end{align}
and we write the summation as $\sum_{(l,m)} := \sum_{l \geq 1} \sum_{|m|\leq l}$.
The KK mass associated with $l$ is given by 
\begin{align}
    m_l \coloneqq \frac{\sqrt{l(l+1)}}{r_0},
\qquad l\ge1.
\end{align}
The relation between the real and complex spherical harmonics is given in Eq.~\eqref{eq:def-of-real-valued-SH}.
The decomposition yields, for each angular momentum $l$, two-dimensional real vector fields $a_{(l,m)a}$ and two towers of real scalars $\varphi_{(l,m)}$ and $\tilde{\varphi}_{(l,m)}$.
The massive Maxwell KK modes transform nontrivially under the SO(3) gauge transformations~\eqref{eq:SO(3)-transformation-y} and \eqref{eq:SO(3)-transformation-B}.
The details of the $\SO(3)$ gauge transformation are discussed in App.~\ref{App:SO(3)-gauge-symmetry}.

Substituting the decomposition~\eqref{eq:4D-gauge-field-A_a-decomposition} and \eqref{eq:4D-gauge-field-A_k-decomposition} and integrating over $S^2$ gives the reduced Maxwell action.
This reduced action inherits a local U(1) gauge symmetry from the four-dimensional theory.
To define the functional integral, we fix the local U(1) gauge symmetry using the Faddeev--Popov (FP) procedure.
Without graviphotons, one may impose a Lorenz-type gauge condition~\cite{Abe:2026sai}.
Because the KK modes transform nontrivially under SO(3), ordinary derivatives acting on the massive modes are replaced by SO(3)-covariant derivatives, and the condition becomes  
\begin{align}
    D_a a^{a(l,m)} - m_l \tilde{\varphi}^{(l,m)} =0 \quad (l\ge1),
    \label{eq:2D-gauge-fixing-condition}
\end{align}
where the $\SO(3)$ covariant derivative is
\begin{align}
    D_a &:= \nabla_a - B_a^I T^{(l)}_I.
    \label{eq:def-of-D_a-varphi}
\end{align}
Here, $\nabla_a$ denotes the Levi-Civita covariant derivative associated with $g_{ab}$.
The graviphotons gauge the SO(3) isometry of $S^2$, so their coupling to the KK modes is determined by the action of the Killing vectors $\zeta_I$ on the spherical harmonics.
Since this action preserves $l$ and mixes only the magnetic quantum number $m$, it can be represented on each $(2l+1)$-dimensional $l$ multiplet by a matrix $T_I^{(l)}$, defined by
\begin{align}
    \sum_{|n|\le l} (T_I^{(l)})_{nm} Y^{\mathbb{R}}_{(l,n)}
    := \zeta^\alpha_I \partial_\alpha Y^{\mathbb{R}}_{(l,m)}.
    \label{eq:def-of-T_I}
\end{align}
For the real spherical harmonics used here, $T_I^{(l)}$ is antisymmetric.
See App.~\ref{App:Representation-matrices} for details.
We do not gauge-fix the massless mode $\delta a_a$, since it is treated as a background field and is not integrated over; only the massive KK modes are integrated out.

The gauge-fixed Maxwell action becomes
\begin{align}
    I_{\text{Maxwell}} + I_{\text{FP+GF}} = I_0 + I_{\text{mixing}},
    \label{eq:gauge-fixed-Maxwell-action-1}
\end{align}
where $I_{\text{FP+GF}}$ is the gauge-fixing and ghost action given in Eq.~\eqref{Eq:FP+GF}, and the two-dimensional actions $I_0$ and $I_{\mathrm{mixing}}$ are
\begin{align}
    I_0 &=
    \frac{r_0^2}{4e^2}\int_\Sigma d^2x \sqrt{\det g}f_{ab} f^{ab}
    \nn \\
    &\quad +
    \frac{r_0^2}{2e^2}\sum_{(l,m)}\int_\Sigma d^2x \sqrt{\det g}
    \Big[
        2 b^{(l,m)}(-D^2 + m_l^2)  c^{(l,m)}
        +\varphi^{(l,m)} (-D^2 + m_l^2)  \varphi^{(l,m)}
        \notag\\
        &\quad
        + \tilde{\varphi}^{(l,m)} (-D^2 + m_l^2)  \tilde{\varphi}^{(l,m)}
        +
        a_a^{(l,m)} (-D^2g^{ab} -[D^a,D^b] + m_l^2 g^{ab})  a_b^{(l,m)}
    \Big],
    \label{eq:I_0}
    \\
    I_{\text{mixing}} &=
    \frac{r_0^2}{e^2}\sum_{(l,m)}\int_\Sigma d^2x \sqrt{\det g}
    \Bigg[
        \frac{1}{2m_l} f^{ab} H_{ab}^I (S_I)_{(0,0)(l,m)} \varphi^{(l,m)}
    \notag\\
    &\quad+
        \sum_{(l',m')}\frac{1}{4m_l m_{l'}}  H_{ab}^I H^{Jab} (S_I)_{(0,0)(l,m)} (S_J)_{(0,0)(l',m')} \varphi^{(l,m)} \varphi^{(l',m')}
    \Bigg]
    \notag\\
    & \quad + \frac{r_0^2}{e^2} \sum_{(l,m)}\int_\Sigma d^2x \sqrt{\det g}
    \Biggl[
        f^{(l,m)}_{ab} \Bigl( \frac{1}{2m_l} H^{Iab} \sum_{|n|\le l}(T_I^{(l)})_{mn} \tilde{\varphi}^{(l,n)} + \sum_{(l',m')} \frac{1}{2m_{l'}} H^{Iab} (S_I)_{(l,m)(l',m')} \varphi^{(l',m')} \Bigr)
        \notag\\
        &\quad
        +\Bigl( \frac{1}{2m_l} H^{Iab} \sum_{|n|\le l}(T_I^{(l)})_{mn} \tilde{\varphi}^{(l,n)} + \sum_{(l',m')} \frac{1}{2m_{l'}} H^{Iab} (S_I)_{(l,m)(l',m')} \varphi^{(l',m')} \Bigr)^2
    \Biggr].
    \label{eq:I_mixing}
\end{align}
See App.~\ref{App:Maxwell-action} for details of the dimensional reduction.
For $l\geq1$, $(b^{(l,m)},c^{(l,m)})$ are FP ghosts, and we define
\begin{align}
    f^{(l,m)}_{ab}:=D_a a_b^{(l,m)}-D_b a_a^{(l,m)}.
\end{align}
On the classical RN black hole background, the massive KK fields have vanishing background values,
while the massless field strength $f_{ab}$ contains both the background electromagnetic field and its fluctuation:
\begin{align}
    f_{ab}
    =
    i\frac{e^2Q}{\sqrt{4\pi}r_0^2}\epsilon_{ab}
    +\delta f_{ab},
    \qquad
    \delta f_{ab}:=\partial_a\delta a_b-\partial_b\delta a_a.
    \label{eq:f-decomposion}
\end{align}
Here, we have used the RN background field strength
\begin{align}
    F_{ab}^{\mathrm{RN}}
    =
    \partial_a A_b^{\mathrm{RN}}
    -\partial_b A_a^{\mathrm{RN}}
    =
    i\frac{e^2Q}{4\pi r_0^2}\epsilon_{ab},
\end{align}
where $\epsilon_{ab}$ denotes the two-dimensional antisymmetric tensor.
The different numerical factors arise from the normalization
$f_{ab}=\sqrt{4\pi}\,F_{ab}$ used in the dimensional reduction.

The graviphoton interactions also involve the action of the Killing vectors on the axial vector spherical harmonics, which can mix KK modes with different angular momenta.
We encode this mixing in the matrix $S_I$, defined by
\begin{align}
    \sum_{l'=0} \sum_{|m'|\le l'}
    (S_I)_{(l',m')(l,m)} Y^{\mathbb{R}}_{(l',m')}
    :=
    \zeta^\alpha_I \epsilon_{\alpha\beta} \partial^\beta
    Y^{\mathbb{R}}_{(l,m)}.
    \label{eq:def-of-S_I}
\end{align}
Examples of $T_I$ and $S_I$ are given in App.~\ref{App:Representation-matrices}.
The Killing vectors and the Clebsch--Gordan decomposition imply that
$(S_I)_{(l',m')(l,m)}$ vanishes unless $l'=l\pm1$; see Eq.~\eqref{eq:S_I-CG}.
Therefore, a single $S_I$ changes the angular momentum of a KK mode by one unit.
The interaction terms proportional to $H_{ab}^I S_I$ in the reduced action~\eqref{eq:gauge-fixed-Maxwell-action-1}
thus connect modes with angular momenta $l$ and $l\pm1$.
Likewise, terms proportional to $H_{ab}^I H^{Jab}S_I S_J$, which contain two such factors,
can connect a mode with angular momentum $l$ to modes with $l$ or $l\pm2$.
These selection rules reflect angular momentum conservation, since each graviphoton originates from an $l=1$ mode on $S^2$.

The action also contains a
$\widetilde\varphi$--$\varphi$ mixing term proportional to
\begin{align}
H_{ab}^I H^{Jab}(T_I^{(l)})_{mn}(S_J)_{(l,m)(l',m')}
\widetilde\varphi^{(l,n)}\varphi^{(l',m')}.
\end{align}
Since $S_J$ is nonzero only for $l'=l\pm1$, this term mixes different KK
levels and is absent in the $l=1$ truncation used below. Moreover, being
already quadratic in the graviphoton background, this off-diagonal mixing
does not contribute to the gauge kinetic correction considered below.

\subsection{Corrections from the $l=1$ massive modes to the massless modes}
\label{sec:Corrections-from-the-massive-modes-to-the-massless-modes}

We integrate out the massive KK modes to obtain the effective action for the massless sector:
\begin{align}
    e^{-I_{\text{massless}}-\Delta W}
    :=
    \int[da_a^{(l,m)}d\varphi^{(l,m)}d\tilde{\varphi}^{(l,m)}
    db^{(l,m)}dc^{(l,m)}]\,
    e^{-I_{\text{EH}}-I_{\text{GHY}}-I_{\text{Maxwell}}-I_{\text{FP+GF}}}.
    \label{eq:path-integral-massive-KK-modes}
\end{align}
Here, $I_{\text{massless}}$ is the massless sector action given in
Eq.~\eqref{eq:I_massless}, while $\Delta W$ denotes the correction generated
by integrating out the massive KK modes.
Accordingly, in the diagrammatic expansion below, only the massive KK fields and the associated
FP ghosts propagate in internal lines, while the massless fields are treated as background or
external fields.
We focus on the terms in $\Delta W$ that correct the $\U(1)$ and $\SO(3)$
gauge kinetic terms:
\begin{align}
    \Delta W \ni \int_{\Sigma} d^2x \sqrt{\det g}
    \left[
        c_{\text{U(1)}} \delta f_{ab} \delta f^{ab}
        + c_{\text{SO(3)}} H^I_{ab} H^{Iab}
    \right].
    \label{eq:Delta_W}
\end{align}
A bilinear kinetic mixing term $\delta f^{ab}H_{ab}^I$ is forbidden by the SO(3) gauge invariance. We retain only curvature-independent gauge-coupling corrections, omitting terms such as $R_{(2)}f_{ab}^2$ and $R_{(2)}H_{ab}^2$.
Within the curvature-independent approximation, $c_{\text{U(1)}}$ and $c_{\text{SO(3)}}$ are independent of $g_{ab}$, so we may set $g_{ab} = \delta_{ab}$ to compute $c_{\text{U(1)}}$ and $c_{\text{SO(3)}}$.

For simplicity, we first restrict the analysis to the lightest massive modes with $l=1$
and discuss possible contributions from $l\geq2$ modes in
Sec.~\ref{sec:Corrections-from-the-l>1-massive-modes}.
In particular, the $l=1$ massive KK modes couple to the
graviphotons through the background electric flux $F_{ab}^{\mathrm{RN}}$.
Integrating out these modes generates a correction to the $\SO(3)$ gauge coupling
proportional to $Q^2$, where $Q$ is the background black hole charge.

Before turning to the perturbative calculation, we note that boundary-condition effects
and the Weyl anomaly associated with the transformation~(2.20) are not included in the
analysis below; we discuss these effects in
Sec.~\ref{sec:Corrections-from-the-l>1-massive-modes}.
We evaluate the coefficients $c_{\text{U(1)}}$ and $c_{\text{SO(3)}}$
perturbatively using Feynman diagrams.
We first derive the relevant Feynman rules and then evaluate the tree-level and
one-loop diagrams contributing to the effective action.
Details of the calculation are presented in
App.~\ref{App:Evaluation-of-effective-action}.

\subsubsection*{Feynman rules}

We use the Feynman rules derived from the reduced Maxwell action~\eqref{eq:gauge-fixed-Maxwell-action-1}. 
Here, the curly and wavy lines denote graviphotons and U(1) photons, respectively. The dashed lines denote the l = 1 massive scalars $\varphi^{(1,I)}$ and $\tilde{\varphi}^{(1,I)}$, and the wavy lines connecting two internal vertices denote the $l = 1$ massive vectors $a_a^{(1,I)}$. The dotted line with an arrow represents the ghosts $b^{(1,I)}$ and $c^{(1,I)}$. 
Crosses $\otimes$ denote insertions of the classical RN black hole background field $F_{ab}^{\mathrm{RN}}$. The circled dot $\circ$ denotes a vertex involving the SO(3) field strength $H_{ab}^I$, whereas the filled dot $\bullet$ denotes a vertex involving the SO(3) gauge field $B_a^I$ directly.

\paragraph{Propagators.}

The propagators of the massive modes and FP ghosts are as follows:
\begin{align}
    \braket{\varphi^{(1,I)}(x) \varphi^{(1,J)}(x')} 
    &=
    \begin{tikzpicture}[baseline={(current bounding box.center)}]
        \begin{feynhand}
            \vertex [] (a) at (-1.5,0) {$\varphi^{(1,I)}(x)$};
            \vertex [] (b) at (1.5,0) {$\varphi^{(1,J)}(x')$};
            \propag [scalar] (a) to (b);
        \end{feynhand}
    \end{tikzpicture}
    = \frac{e^2}{r_0^2} \frac{\delta_{IJ}}{-\del^2_x + m_1^2} \delta(x-x'),
    \\
    \braket{\tilde{\varphi}^{(1,I)}(x) \tilde{\varphi}^{(1,J)}(x')} 
    &=
    \begin{tikzpicture}[baseline={(current bounding box.center)}]
        \begin{feynhand}
            \vertex [] (a) at (-1.5,0) {$\tilde{\varphi}^{(1,I)}(x)$};
            \vertex [] (b) at (1.5,0) {$\tilde{\varphi}^{(1,J)}(x')$};
            \propag [scalar] (a) to (b);
        \end{feynhand}
    \end{tikzpicture}
    = \frac{e^2}{r_0^2} \frac{\delta_{IJ}}{-\del^2_x + m_1^2} \delta(x-x'),
    \\
    \braket{a_a^{(1,I)}(x) a_b^{(1,J)}(x')} 
    &=
    \begin{tikzpicture}[baseline={(current bounding box.center)}]
        \begin{feynhand}
            \vertex [] (a) at (-1.5,0) {$a_a^{(1,I)}(x)$};
            \vertex [] (b) at (1.5,0) {$a_b^{(1,J)}(x')$};
            \propag [boson] (a) to (b);
        \end{feynhand}
    \end{tikzpicture}
    = \frac{e^2}{r_0^2} \frac{\delta_{IJ}\delta_{ab}}{-\del^2_x + m_1^2} \delta(x-x'),
    \\
    \braket{c^{(1,I)}(x) b^{(1,J)}(x')} 
    &=
    \begin{tikzpicture}[baseline={(current bounding box.center)}]
        \begin{feynhand}
            \vertex [] (a) at (-1.5,0) {$c^{(1,I)}(x)$};
            \vertex [] (b) at (1.5,0) {$b^{(1,J)}(x')$};
            \propag [chagho] (a) to (b);
        \end{feynhand}
    \end{tikzpicture}
    = \frac{e^2}{r_0^2} \frac{\delta_{IJ}}{-\del^2_x + m_1^2} \delta(x-x').
\end{align}

\paragraph{Interactions arising from the mode mixing terms.}
Interactions arising from the mode-mixing terms~\eqref{eq:I_mixing} are as follows:
\begin{itemize}
    \item $f^{ab}H_{ab}^I (S_I)_{(0,0)(1,J)}\varphi^{(1,J)}$ term:
        Splitting $f_{ab}$ into the classical background $F_{ab}^{\text{RN}}$ and quantum fluctuation $\delta f_{ab}$ by using Eq.~\eqref{eq:f-decomposion}, we obtain
            \begin{align}
                \begin{tikzpicture}[baseline={(current bounding box.center)}]
                    \begin{feynhand}
                        \vertex [inner sep=0pt, label=above:$F^{\text{RN}}_{ab}$] (a) at (0,1) {$\otimes$};
                        \vertex [label=left:$H_{ab}^I$] (b) at (-1.5,0) {};
                        \vertex [label=right:$\varphi^{(1,J)}$] (c) at (1.5,0) {};
                        \vertex [ringdot] (v) at (0,0) {};
                        \propag [pho] (a) to (v);
                        \propag [glu] (b) to (v);
                        \propag [scalar] (v) to (c);
                    \end{feynhand}
                \end{tikzpicture}
                &= -\frac{r_0^2}{e^2} \frac{1}{2m_1} \sqrt{4\pi} F^{\text{RN} ab}(S_I)_{(0,0)(1,J)} = - i \frac{r_0 Q}{\sqrt{24\pi}} \epsilon^{ab} \delta_{IJ},
                \label{eq:H-deltaf-varphi-vartex}
                \\
                \begin{tikzpicture}[baseline={(current bounding box.center)}]
                    \begin{feynhand}
                        \vertex [label=left:$\delta f_{cd}$] (a) at (-1.5,1) {};
                        \vertex [label=left:$H_{ab}^I$] (b) at (-1.5,-1) {};
                        \vertex [label=right:$\varphi^{(1,J)}$] (c) at (1.5,0) {};
                        \vertex [ringdot] (v) at (0,0) {};
                        \propag [pho] (a) to (v);
                        \propag [glu] (b) to (v);
                        \propag [scalar] (v) to (c);
                    \end{feynhand}
                \end{tikzpicture}
                &= 
                -\frac{r_0^2}{e^2} \frac{1}{4m_1} \left(\delta^{ac}\delta^{bd} - \delta^{ad}\delta^{bc} \right) (S_I)_{(0,0)(1,J)}
                \notag\\
                &=
                - \frac{r_0^3}{2\sqrt{6} e^2}\left(\delta^{ac}\delta^{bd} - \delta^{ad}\delta^{bc} \right) \delta_{IJ},
                \label{eq:H-f-varphi-vartex}
            \end{align}
            where we have used 
            \begin{align}
                (S_I)_{(0,0)(1,J)} = \frac{2}{\sqrt{3}} \delta_{IJ},
            \end{align}
            whose derivation is given in Eq.~\eqref{eq:S_I_00_1J}.
    \item $f^{Kab}H_{ab}^I (T^{(1)}_I)_{KJ}\tilde{\varphi}^{(1,J)}$ term: 
            Using $f^{Kab}H_{ab}^I= 2(\partial^a a^{Kb})H_{ab}^I - 2B^{I'a} (T_{I'}^{(1)})_{KJ} a^{Jb} H_{ab}^I$, this term gives rise to the following two vertices:
            \begin{align}
                \begin{tikzpicture}[baseline={(current bounding box.center)}]
                    \begin{feynhand}
                        \vertex [label=left:$a_c^{(1,J)}$] (a) at (-1.5,1) {};
                        \vertex [label=right:$\tilde{\varphi}^{(1,K)}$] (b) at (1.5,0) {};
                        \vertex [label=left:$H_{ab}^I$] (c) at (-1.5,-1) {};
                        \vertex [ringdot] (v) at (0,0) {};
                        \propag [boson, mom={[arrow style] $p$}] (a) to (v);
                        \propag [scalar, mom={[arrow style] $k+p$}] (v) to (b);
                        \propag [glu, mom={[arrow style] $k$}] (c) to (v);
                    \end{feynhand}
                \end{tikzpicture}
                &=
                \frac{
                r_0^2}{e^2}\frac{-i}{2m_1} \left( p^a \delta^{bc} - p^b \delta^{ac} \right) (T_I^{(1)})_{JK},
                \\
                \begin{tikzpicture}[baseline={(current bounding box.center)}]
                    \begin{feynhand}
                        \vertex [label=left:$H_{ab}^{I'}$] (a) at (-1.5,1) {};
                        \vertex [label=left:$B_{c}^{J'}$] (b) at (-1.5,-1) {};
                        \vertex [label=right:$a_d^{(1,I)}$] (c) at (1.5,1) {};
                        \vertex [label=right:$\tilde{\varphi}^{(1,J)}$] (d) at (1.5,-1) {};
                        \vertex [ringdot] (v) at (0,0) {};
                        \propag [glu] (a) to (v);
                        \propag [glu] (b) to (v);
                        \propag [boson] (v) to (c);
                        \propag [scalar] (v) to (d);
                    \end{feynhand}
                \end{tikzpicture}
                &= 
                    -\frac{r_0^2}{e^2}\frac{1}{2m_1} \left(\delta^{ac}\delta^{bd}- \delta^{ad}\delta^{bc}\right)  \sum_{K=0,\pm 1}(T_{I'}^{(1)})_{JK} (T_{J'}^{(1)})_{KI}.
            \end{align}
    \item $H_{ab}^{I'} H^{J'ab} (S_{I'})_{(l,m)(1,I)} (S_{J'})_{(l,m)(1,J)} \varphi^{(1,I)} \varphi^{(1,J)}$ term:
            \begin{align}
                &\begin{tikzpicture}[baseline={(current bounding box.center)}]
                    \begin{feynhand}
                        \vertex [label=left:$H_{ab}^{I'}$] (a) at (-1.5,1) {};
                        \vertex [label=left:$H_{cd}^{J'}$] (b) at (-1.5,-1) {};
                        \vertex [label=right:$\varphi^{(1,I)}$] (c) at (1.5,1) {};
                        \vertex [label=right:$\varphi^{(1,J)}$] (d) at (1.5,-1) {};
                        \vertex [ringdot] (v) at (0,0) {};
                        \propag [glu] (a) to (v);
                        \propag [glu] (b) to (v);
                        \propag [scalar] (v) to (c);
                        \propag [scalar] (v) to (d);
                    \end{feynhand}
                \end{tikzpicture}
                \notag\\
                &= - \frac{r_0^2}{e^2}\frac{1}{4 m_1^2} \sum_{l=0}^{2} \sum_{|m| \leq l} \left(\delta^{ac} \delta^{bd} - \delta^{ad} \delta^{bc} \right) \left(
                    (S_{I'})_{(l,m)(1,I)} (S_{J'})_{(l,m)(1,J)} + (S_{I'})_{(l,m)(1,J)} (S_{J'})_{(l,m)(1,I)}
                \right).
                \label{eq:mode-mixing-HH-varphi-varphi}
            \end{align}
            
    \item $H_{ab}^{I'} H^{J'ab} (T^{(1)}_{I'})_{KI} (T^{(1)}_{J'})_{KJ} \tilde{\varphi}^{(1,I)} \tilde{\varphi}^{(1,J)}$ term:
            \begin{align}
                &\begin{tikzpicture}[baseline={(current bounding box.center)}]
                    \begin{feynhand}
                        \vertex [label=left:$H_{ab}^{I'}$] (a) at (-1.5,1) {};
                        \vertex [label=left:$H_{cd}^{J'}$] (b) at (-1.5,-1) {};
                        \vertex [label=right:$\tilde{\varphi}^{(1,I)}$] (c) at (1.5,1) {};
                        \vertex [label=right:$\tilde{\varphi}^{(1,J)}$] (d) at (1.5,-1) {};
                        \vertex [ringdot] (v) at (0,0) {};
                        \propag [glu] (a) to (v);
                        \propag [glu] (b) to (v);
                        \propag [scalar] (v) to (c);
                        \propag [scalar] (v) to (d);
                    \end{feynhand}
                \end{tikzpicture}
                \notag\\
                &= - \frac{r_0^2}{e^2}\frac{1}{4 m_1^2} \left(\delta^{ac} \delta^{bd} - \delta^{ad} \delta^{bc} \right) \sum_{K =0, \pm1} \left(
                    (T^{(1)}_{I'})_{KI} (T^{(1)}_{J'})_{KJ} + (T^{(1)}_{I'})_{KJ} (T^{(1)}_{J'})_{KI}
                \right).
                \label{eq:mode-mixing-HH-tildevarphi-tildevarphi}
            \end{align}
            
\end{itemize}

\paragraph{Interactions arising from the covariant derivative terms.}
Interactions arising from the covariant derivative terms in Eq.~\eqref{eq:I_0} are as follows:
\begin{itemize}
    \item  The three-point interactions between one graviphoton  $B_a^I$ and two $l=1$ KK modes:
        \begin{align}
            \begin{tikzpicture}[baseline={(current bounding box.center)}]
                \begin{feynhand}
                    \vertex [label=left:$\varphi^{(1,J)}$] (a) at (-1.5,1) {};
                    \vertex [label=right:$\varphi^{(1,K)}$] (b) at (1.5,0) {};
                    \vertex [label=left:$B_a^I$] (c) at (-1.5,-1) {};
                    \vertex [dot] (v) at (0,0) {};
                    \propag [scalar, mom={[arrow style] $p$}] (a) to (v);
                    \propag [scalar, mom={[arrow style] $k+p$}] (v) to (b);
                    \propag [glu, mom={[arrow style] $k$}] (c) to (v);
                \end{feynhand}
            \end{tikzpicture}
            &= i\frac{r_0^2}{e^2}(2p^a + k^a) (T_I^{(1)})_{JK},
            \label{eq:B_varphi_vaprhi}
            \\
            \begin{tikzpicture}[baseline={(current bounding box.center)}]
                \begin{feynhand}
                    \vertex [label=left:$\tilde{\varphi}^{(1,J)}$] (a) at (-1.5,1) {};
                    \vertex [label=right:$\tilde{\varphi}^{(1,K)}$] (b) at (1.5,0) {};
                    \vertex [label=left:$B_a^I$] (c) at (-1.5,-1) {};
                    \vertex [dot] (v) at (0,0) {};
                    \propag [plain, mom={[arrow style] $p$}] (a) to (v);
                    \propag [plain, mom={[arrow style] $k+p$}] (v) to (b);
                    \propag [glu, mom={[arrow style] $k$}] (c) to (v);
                \end{feynhand}
            \end{tikzpicture}
            &= i\frac{r_0^2}{e^2}(2p^a + k^a) (T_I^{(1)})_{JK},
            \label{eq:B_tilde-varphi_tilde-vaprhi}
            \\
            \begin{tikzpicture}[baseline={(current bounding box.center)}]
                \begin{feynhand}
                    \vertex [label=left:$b^{(1,J)}$] (a) at (-1.5,1) {};
                    \vertex [label=right:$c^{(1,K)}$] (b) at (1.5,0) {};
                    \vertex [label=left:$B_a^I$] (c) at (-1.5,-1) {};
                    \vertex [dot] (v) at (0,0) {};
                    \propag [ghost, mom={[arrow style] $p$}] (a) to (v);
                    \propag [ghost, mom={[arrow style] $k+p$}] (v) to (b);
                    \propag [glu, mom={[arrow style] $k$}] (c) to (v);
                \end{feynhand}
            \end{tikzpicture}
            &= i\frac{r_0^2}{e^2}(2p^a + k^a) (T_I^{(1)})_{JK},
            \label{eq:B_b_c}
            \\
            \begin{tikzpicture}[baseline={(current bounding box.center)}]
                \begin{feynhand}
                    \vertex [label=left:$a_c^{(1,J)}$] (a) at (-1.5,1) {};
                    \vertex [label=right:$a_d^{(1,K)}$] (b) at (1.5,0) {};
                    \vertex [label=left:$B_a^I$] (c) at (-1.5,-1) {};
                    \vertex [dot] (v) at (0,0) {};
                    \propag [boson, mom={[arrow style] $p$}] (a) to (v);
                    \propag [boson, mom={[arrow style] $k+p$}] (v) to (b);
                    \propag [glu, mom={[arrow style] $k$}] (c) to (v);
                \end{feynhand}
            \end{tikzpicture}
            &= i\frac{r_0^2}{e^2}\delta^{cd}(2p^a + k^a) (T_I^{(1)})_{JK}.
            \label{eq:B_a_a}
        \end{align}
    \item   The four-point interactions between two graviphotons $B_a^{I}$ and two $l=1$ KK modes:
            \begin{align}
                \begin{tikzpicture}[baseline={(current bounding box.center)}]
                    \begin{feynhand}
                        \vertex [label=left:$B_{a}^{I'}$] (a) at (-1.5,1) {};
                        \vertex [label=left:$B_{b}^{J'}$] (b) at (-1.5,-1) {};
                        \vertex [label=right:$\varphi^{(1,I)}$] (c) at (1.5,1) {};
                        \vertex [label=right:$\varphi^{(1,J)}$] (d) at (1.5,-1) {};
                        \vertex [dot] (v) at (0,0) {};
                        \propag [glu] (a) to (v);
                        \propag [glu] (b) to (v);
                        \propag [scalar] (v) to (c);
                        \propag [scalar] (v) to (d);
                    \end{feynhand}
                \end{tikzpicture}
                &= -\frac{r_0^2}{e^2}\delta^{ab} \sum_{K=0,\pm1} \left(
                    (T^{(1)}_{I'})_{KI} (T^{(1)}_{J'})_{KJ} + (T^{(1)}_{I'})_{KJ} (T^{(1)}_{J'})_{KI}
                \right),
                \label{eq:B_B_varphi_varphi}
                \\
                \begin{tikzpicture}[baseline={(current bounding box.center)}]
                    \begin{feynhand}
                        \vertex [label=left:$B_{a}^{I'}$] (a) at (-1.5,1) {};
                        \vertex [label=left:$B_{b}^{J'}$] (b) at (-1.5,-1) {};
                        \vertex [label=right:$\tilde{\varphi}^{(1,I)}$] (c) at (1.5,1) {};
                        \vertex [label=right:$\tilde{\varphi}^{(1,J)}$] (d) at (1.5,-1) {};
                        \vertex [dot] (v) at (0,0) {};
                        \propag [glu] (a) to (v);
                        \propag [glu] (b) to (v);
                        \propag [scalar] (v) to (c);
                        \propag [scalar] (v) to (d);
                    \end{feynhand}
                \end{tikzpicture}
                &=
                    -\frac{r_0^2}{e^2}\delta^{ab} \sum_{K=0,\pm1} \left(
                    (T^{(1)}_{I'})_{KI} (T^{(1)}_{J'})_{KJ} + (T^{(1)}_{I'})_{KJ} (T^{(1)}_{J'})_{KI}
                \right),
                \\
                \begin{tikzpicture}[baseline={(current bounding box.center)}]
                    \begin{feynhand}
                        \vertex [label=left:$B_{a}^{I'}$] (a) at (-1.5,1) {};
                        \vertex [label=left:$B_{b}^{J'}$] (b) at (-1.5,-1) {};
                        \vertex [label=right:$c^{(1,I)}$] (c) at (1.5,1) {};
                        \vertex [label=right:$b^{(1,J)}$] (d) at (1.5,-1) {};
                        \vertex [dot] (v) at (0,0) {};
                        \propag [glu] (a) to (v);
                        \propag [glu] (b) to (v);
                        \propag [ghost] (v) to (c);
                        \propag [ghost] (v) to (d);
                    \end{feynhand}
                \end{tikzpicture}
                &=
                    -\frac{r_0^2}{e^2}\delta^{ab} \sum_{K=0,\pm1} \left(
                    (T^{(1)}_{I'})_{KI} (T^{(1)}_{J'})_{KJ} + (T^{(1)}_{I'})_{KJ} (T^{(1)}_{J'})_{KI}
                \right),
                \label{eq:B_B_b_c}
                \\
                \begin{tikzpicture}[baseline={(current bounding box.center)}]
                    \begin{feynhand}
                        \vertex [label=left:$B_{a}^{I'}$] (a) at (-1.5,1) {};
                        \vertex [label=left:$B_{b}^{J'}$] (b) at (-1.5,-1) {};
                        \vertex [label=right:$a_c^{(1,I)}$] (c) at (1.5,1) {};
                        \vertex [label=right:$a_d^{(1,J)}$] (d) at (1.5,-1) {};
                        \vertex [dot] (v) at (0,0) {};
                        \propag [glu] (a) to (v);
                        \propag [glu] (b) to (v);
                        \propag [boson] (v) to (c);
                        \propag [boson] (v) to (d);
                    \end{feynhand}
                \end{tikzpicture}
                &=
                    -\frac{r_0^2}{e^2}\delta^{ab} \delta^{cd} \sum_{K=0,\pm1} \left(
                    (T^{(1)}_{I'})_{KI} (T^{(1)}_{J'})_{KJ} + (T^{(1)}_{I'})_{KJ} (T^{(1)}_{J'})_{KI}
                \right).
                \label{eq:B_B_a_a}
            \end{align}
            
        \item   The three-point interactions between one graviphoton  $H_{ab}^I$ and two $l=1$ KK modes $a_a^{(1,I)}$:
                \begin{align}
                    \begin{tikzpicture}[baseline={(current bounding box.center)}]
                        \begin{feynhand}
                            \vertex [label=right:$a_c^{(1,J)}$] (a) at (1.5,1) {};
                            \vertex [label=right:$a_d^{(1,K)}$] (b) at (1.5,-1) {};
                            \vertex [label=left:$H_{ab}^I$] (c) at (-1.5,0) {};
                            \vertex [ringdot] (v) at (0,0) {};
                            \propag [boson] (a) to (v);
                            \propag [boson] (b) to (v);
                            \propag [glu] (v) to (c);
                        \end{feynhand}
                    \end{tikzpicture}
                    = \frac{r_0^2}{2e^2} \left( \delta^{ac}\delta^{bd} - \delta^{ad}\delta^{bc} \right)(T_I^{(1)})_{KJ}.
                \end{align}
\end{itemize}

\subsubsection*{Corrections from massive KK modes}

Using the Feynman rules above, we perturbatively evaluate the massive KK mode contributions to the effective action, focusing on the U(1) and SO(3) gauge couplings.
At the order considered here, the two-dimensional photon has no direct coupling to the massive KK modes in the absence of graviphotons.
Thus, the $l=1$ massive modes generate no terms proportional to $\delta f_{ab}\delta f^{ab}$ at quadratic order in the massless fluctuations, so
\begin{align}
    c_{\mathrm{U}(1)}\approx 0,
    \label{eq:c-U1-0}
\end{align}
within the approximation considered here. 
The $l \ge 2$ modes likewise do not contribute to $c_{\text{U(1)}}$.

At quadratic order in the massless fluctuations and at leading order in the
derivative expansion, the contributions to $c_{\mathrm{SO}(3)}$ arise at
both tree level and one loop.
Within the curvature-independent approximation stated above, we assume that
the characteristic external graviphoton momentum satisfies
\begin{align}
    \frac{k^2}{m_1^2} \ll 1.
\end{align}
This condition controls only the momentum expansion; curvature-dependent
terms, with relative size $O(R_{(2)}/m_1^2)$, are not evaluated here.
We therefore neglect relative corrections of order $k^2/m_1^2$.
Details of the calculation are presented in App.~\ref{App:Evaluation-of-effective-action}.

\begin{itemize}
\item \textbf{Tree contribution}:
This contribution is generated by exchange of
$\varphi^{(1,I)}$. 
Since each vertex contains one insertion of the RN black hole background $F^{\text{RN}}_{ab} \propto Q\epsilon_{ab}$, the resulting correction is proportional
to $Q^2$:
\begin{align}
    \begin{tikzpicture}[baseline={(current bounding box.center)}]
    \begin{feynhand}
        \vertex [inner sep=0pt, label=above:$F^{\text{RN}}_{ab}$] (a) at (0,1) {$\otimes$};
        \vertex [label=left:$H_{ab}^I$] (b) at (-1,0) {};
        \vertex [ringdot] (v) at (0,0) {};
        \vertex [inner sep=0pt, label=above:$F^{\text{RN}}_{cd}$] (a2) at (1.5,1) {$\otimes$};
        \vertex [label=right:$H_{cd}^I$] (b2) at (2.5,0) {};
        \vertex [ringdot] (v2) at (1.5,0) {};
        \propag [pho] (a) to (v);
        \propag [glu] (b) to (v);
        \propag [pho] (a2) to (v2);
        \propag [glu] (b2) to (v2);
        \propag [scalar] (v) to [edge label' = $\varphi^{(1,J)}$] (v2);
    \end{feynhand}
    \end{tikzpicture}
    &= \frac{e^2 Q^2}{24\pi} \int_\Sigma d^2x \sqrt{\det g}   ~H^I_{ab} (-D^2 + m_1^2)^{-1} H^{Iab}
    \label{eq:tree-level-correction-from-l=1w/propagator}
    \\
    &\approx \frac{r_0^2 e^2 Q^2}{48\pi} \int_\Sigma d^2x \sqrt{\det g}   ~H^I_{ab} H^{Iab},
    \label{eq:tree-level-correction-from-l=1}
\end{align}
Here, $D_a$ is the SO(3)-covariant derivative in the $l = 1$ representation, and $m_1^2 =2/r_0^2$.
We have restored the covariant measure $\sqrt{\det g} $ here and do so below whenever presenting covariant expressions.
At quadratic order in the graviphoton background, its $B_I^a$-dependent part may be neglected inside the propagator, 
reproducing the free propagator used in the diagrammatic calculation.
\item \textbf{One-loop contributions}:
Massive KK modes and FP ghosts generate the one-loop contribution.
The correction to the effective action is evaluated as 
\begin{align}
    &
    \begin{tikzpicture}[baseline={(current bounding box.center)}]
    \begin{feynhand}
        \vertex [](a) at (-1.5,0) {};
        \vertex [dot] (v) at (0,0) {};
        \vertex [] (b) at (1.5,0) {};
        \propag [glu] (a) to (v);
        \propag [glu] (v) to (b);
        \draw [scalar] (v) arc (-90:270:0.5) node[midway, above] {$X^{(1,I)}$};
    \end{feynhand}
    \end{tikzpicture}
    +
    \begin{tikzpicture}[baseline={(current bounding box.center)}]
    \begin{feynhand}
        \vertex [](a) at (-1.5,0) {};
        \vertex [ringdot] (v) at (0,0) {};
        \vertex [] (b) at (1.5,0) {};
        \propag [glu] (a) to (v);
        \propag [glu] (v) to (b);
        \draw [scalar] (v) arc (-90:270:0.5) node[midway, above] {$X^{(1,I)}$};
    \end{feynhand}
    \end{tikzpicture}
    +
    \begin{tikzpicture}[baseline={(current bounding box.center)}]
    \begin{feynhand}
        \vertex [] (a) at (-1.5,0) {};
        \vertex [] (b) at (1.5,0) {};
        \vertex [dot] (v1) at (-0.5,0) {};
        \vertex [dot] (v2) at (0.5,0) {};
        \propag [glu] (a) to (v1);
        \propag [glu] (v2) to (b);
        \draw [scalar] (v1) arc (180:0:0.5) node[midway, above] {$X^{(1,I)}$};
        \draw [scalar] (v2) arc (0:-180:0.5) node[midway, below] {$X^{(1,J)}$};
    \end{feynhand}
    \end{tikzpicture}
    +
    \begin{tikzpicture}[baseline={(current bounding box.center)}]
    \begin{feynhand}
        \vertex [] (a) at (-1.5,0) {};
        \vertex [] (b) at (1.5,0) {};
        \vertex [ringdot] (v1) at (-0.5,0) {};
        \vertex [ringdot] (v2) at (0.5,0) {};
        \propag [glu] (a) to (v1);
        \propag [glu] (v2) to (b);
        \draw [scalar] (v1) arc (180:0:0.5) node[midway, above] {$X^{(1,I)}$};
        \draw [scalar] (v2) arc (0:-180:0.5) node[midway, below] {$X^{(1,J)}$};
    \end{feynhand}
    \end{tikzpicture}
    \notag\\
    &
    \begin{tikzpicture}[baseline={(current bounding box.center)}]
    \begin{feynhand}
        \vertex [](a) at (-1.5,0) {};
        \vertex [dot] (v) at (0,0) {};
        \vertex [] (b) at (1.5,0) {};
        \propag [glu] (a) to (v);
        \propag [glu] (v) to (b);
        \draw [boson] (v) arc (-90:270:0.5) node[midway, above] {$a_a^{(1,I)}$};
    \end{feynhand}
    \end{tikzpicture}
    +
    \begin{tikzpicture}[baseline={(current bounding box.center)}]
    \begin{feynhand}
        \vertex [](a) at (-1.5,0) {};
        \vertex [ringdot] (v) at (0,0) {};
        \vertex [] (b) at (1.5,0) {};
        \propag [glu] (a) to (v);
        \propag [glu] (v) to (b);
        \draw [boson] (v) arc (-90:270:0.5) node[midway, above] {$a_a^{(1,I)}$};
    \end{feynhand}
    \end{tikzpicture}
    +
    \begin{tikzpicture}[baseline={(current bounding box.center)}]
    \begin{feynhand}
        \vertex [] (a) at (-1.5,0) {};
        \vertex [] (b) at (1.5,0) {};
        \vertex [dot] (v1) at (-0.5,0) {};
        \vertex [dot] (v2) at (0.5,0) {};
        \propag [glu] (a) to (v1);
        \propag [glu] (v2) to (b);
        \draw [boson] (v1) arc (180:0:0.5) node[midway, above] {$a_a^{(1,I)}$};
        \draw [boson] (v2) arc (0:-180:0.5) node[midway, below] {$a_b^{(1,J)}$};
    \end{feynhand}
    \end{tikzpicture}
    +
    \begin{tikzpicture}[baseline={(current bounding box.center)}]
    \begin{feynhand}
        \vertex [] (a) at (-1.5,0) {};
        \vertex [] (b) at (1.5,0) {};
        \vertex [ringdot] (v1) at (-0.5,0) {};
        \vertex [ringdot] (v2) at (0.5,0) {};
        \propag [glu] (a) to (v1);
        \propag [glu] (v2) to (b);
        \draw [boson] (v1) arc (180:0:0.5) node[midway, above] {$a_a^{(1,I)}$};
        \draw [boson] (v2) arc (0:-180:0.5) node[midway, below] {$a_b^{(1,J)}$};
    \end{feynhand}
    \end{tikzpicture}
    \notag\\
    &
    +
    \begin{tikzpicture}[baseline={(current bounding box.center)}]
    \begin{feynhand}
        \vertex [] (a) at (-1.5,0) {};
        \vertex [] (b) at (1.5,0) {};
        \vertex [ringdot] (v1) at (-0.5,0) {};
        \vertex [ringdot] (v2) at (0.5,0) {};
        \propag [glu] (a) to (v1);
        \propag [glu] (v2) to (b);
        \draw [scalar] (v1) arc (180:0:0.5) node[midway, above] {$\tilde{\varphi}^{(1,J)}$};
        \draw [boson] (v2) arc (0:-180:0.5) node[midway, below] {$a_a^{(1,J)}$};
    \end{feynhand}
    \end{tikzpicture}
    +
    \begin{tikzpicture}[baseline={(current bounding box.center)}]
    \begin{feynhand}
        \vertex [] (a) at (-1.5,0) {};
        \vertex [] (b) at (1.5,0) {};
        \vertex [dot] (v1) at (-0.5,0) {};
        \vertex [dot] (v2) at (0.5,0) {};
        \propag [glu] (a) to (v1);
        \propag [glu] (v2) to (b);
        \draw [chagho] (v1) arc (180:0:0.5) node[midway, above] {};
        \draw [chagho] (v2) arc (0:-180:0.5) node[midway, below] {};
    \end{feynhand}
    \end{tikzpicture}
    +
    \begin{tikzpicture}[baseline={(current bounding box.center)}]
    \begin{feynhand}
        \vertex [](a) at (-1.5,0) {};
        \vertex [dot] (v) at (0,0) {};
        \vertex [] (b) at (1.5,0) {};
        \propag [glu] (a) to (v);
        \propag [glu] (v) to (b);
        \draw [chagho] (v) arc (-90:270:0.5) node[midway, above] {};
    \end{feynhand}
    \end{tikzpicture}
    \notag\\
    &\approx \frac{r_0^2}{48\pi}\int_\Sigma d^2x \sqrt{\det g}  H^I_{ab} H^{Iab}.
    \label{eq:1-loop-level-correction-from-l=1}
\end{align}
Here, $X^{(1,I)}=\varphi^{(1,I)}$ and $\tilde{\varphi}^{(1,I)}$. 
We recall that the circled dot $\circ$ denotes a vertex involving the SO(3) field strength $H_{ab}^I$, whereas the filled dot $\bullet$ denotes a vertex involving the SO(3) gauge field $B_a^I$ directly.
We use zeta-function regularization and set the renormalization scale to $m_1$. 
\end{itemize}
Combining Eqs.~\eqref{eq:tree-level-correction-from-l=1} and~\eqref{eq:1-loop-level-correction-from-l=1}, we obtain
\begin{align}
    c_{\mathrm{SO(3)}} \approx \frac{r_0^2 e^2 Q^2}{48\pi} + \frac{r_0^2}{48\pi}.
\end{align}
The tree-level contribution is independent of the renormalization prescription, whereas the finite one-loop contribution is scheme dependent.

We now ask under what conditions the corrections induced by integrating out the $l=1$ massive modes remain small relative to the classical graviphoton kinetic coefficient in Eq.~\eqref{eq:I_massless}.
As a necessary criterion, both corrections in Eqs.~\eqref{eq:tree-level-correction-from-l=1} and \eqref{eq:1-loop-level-correction-from-l=1} must be much smaller than the classical coefficient in Eq.~\eqref{eq:I_massless}.
The comparison gives
\begin{align}
    \frac{r_0^2  \Phi(x)}{24G_N} \gg \frac{r_0^2}{48\pi},
    \quad
    \frac{r_0^2  \Phi(x)}{24G_N} \gg \frac{r_0^2 e^2 Q^2}{48\pi}.
    \label{eq:truncation-condition-Phi}
\end{align}
The inequalities in Eq.~\eqref{eq:truncation-condition-Phi} should hold throughout the two-dimensional Euclidean spacetime $\Sigma$.
On the classical RN black hole background, setting $\delta\Phi=0$ gives
$\Phi=r^2\geq r_H^2$, so the conditions~\eqref{eq:truncation-condition-Phi}
are most restrictive at the horizon.
Evaluating Eq.~\eqref{eq:truncation-condition-Phi} at $r=r_H$ gives
\begin{align}
    \left(\frac{r_H}{l_{\text{pl}}}\right)^2 \gg \frac{1}{2\pi},
    \qquad
    \left(\frac{r_H}{l_{\text{pl}}}\right)^2 \gg \frac{e^2Q^2}{2\pi},
    \label{eq:truncation-condition-from-l=1}
\end{align}
where $l_{\text{pl}}:=\sqrt{G_N}$ is the four-dimensional Planck length.
The first inequality is the usual semiclassical condition that the horizon radius is much larger than the Planck length.
The second inequality requires the charge-dependent massive KK correction to remain smaller than the classical graviphoton kinetic term.
Using Eq.~\eqref{eq:r_H&r_pm}, the second condition in Eq.~\eqref{eq:truncation-condition-from-l=1} is parametrically equivalent to
\begin{align}
    G_N M^2\gg \frac{e^2 Q^2}{4\pi}.
    \label{eq:condition-M-Q}
\end{align}
Thus, Eq.~\eqref{eq:condition-M-Q} selects a regime parametrically away from extremality.
In the near-extremal regime defined in Sec.~\ref{sec:4D-charged-black-holes-and-2D-reduced-theory}, Eq.~\eqref{eq:condition-M-Q} is no longer satisfied.
Consequently, within our approximation, the tree-level correction from the $l=1$ massive modes becomes parametrically comparable to the classical graviphoton kinetic term in Eq.~\eqref{eq:I_massless}.

\subsection{More on the corrections from the massive modes}
\label{sec:Corrections-from-the-l>1-massive-modes}

We have so far omitted the $l\geq2$ KK modes and now estimate their contributions.
According to Eq.~\eqref{eq:I_mixing}, these modes do not couple to the background flux $F^{\mathrm{RN}}_{ab}$.
This follows from angular momentum conservation: the RN background flux has $l=0$, while the graviphoton carries $l=1$, so the Clebsch--Gordan selection rule allows a linear coupling only to an $l=1$ massive KK mode.
Therefore, the $l\geq2$ modes generate no $Q$-dependent tree-level correction, although they can contribute $Q$-independent one-loop corrections proportional to $r_0^2$.
The tree-level result~\eqref{eq:tree-level-correction-from-l=1} is thus unchanged.
To estimate the higher-mode one-loop correction, we omit the mixing term~\eqref{eq:I_mixing} for simplicity and consider the contributions from the covariant derivative terms.
For example, the interactions between the graviphotons and $\varphi^{(l,m)}$ are
\begin{align}
    \begin{tikzpicture}[baseline={(current bounding box.center)}]
        \begin{feynhand}
            \vertex [label=left:$\varphi^{(l,n)}$] (a) at (-1.5,1) {};
            \vertex [label=right:$\varphi^{(l,m)}$] (b) at (1.5,0) {};
            \vertex [label=left:$B_a^I$] (c) at (-1.5,-1) {};
            \vertex [dot] (v) at (0,0) {};
            \propag [scalar, mom={[arrow style] $p$}] (a) to (v);
            \propag [scalar, mom={[arrow style] $k+p$}] (v) to (b);
            \propag [glu, mom={[arrow style] $k$}] (c) to (v);
        \end{feynhand}
    \end{tikzpicture}
    & = i\frac{r_0^2}{e^2}(2p^a + k^a) (T_I^{(l)})_{nm},
    \\
    \begin{tikzpicture}[baseline={(current bounding box.center)}]
        \begin{feynhand}
            \vertex [label=left:$B_{a}^{I}$] (a) at (-1.5,1) {};
            \vertex [label=left:$B_{b}^{J}$] (b) at (-1.5,-1) {};
            \vertex [label=right:$\varphi^{(l,m)}$] (c) at (1.5,1) {};
            \vertex [label=right:$\varphi^{(l,n)}$] (d) at (1.5,-1) {};
            \vertex [dot] (v) at (0,0) {};
            \propag [glu] (a) to (v);
            \propag [glu] (b) to (v);
            \propag [scalar] (v) to (c);
            \propag [scalar] (v) to (d);
        \end{feynhand}
    \end{tikzpicture}
    &= -\frac{r_0^2}{e^2}\delta^{ab}\sum_{|n'|\leq l}\left[(T_I^{(l)})_{n'n}(T_J^{(l)})_{n'm}+(T_I^{(l)})_{n'm}(T_J^{(l)})_{n'n}\right].
\end{align}
Following the same calculation as for the $l=1$ modes (App.~\ref{app:One-loop-determinants}), the one-loop contribution from $\varphi^{(l,m)}$ becomes
\begin{align}
    \begin{tikzpicture}[baseline={(current bounding box.center)}]
    \begin{feynhand}
        \vertex [](a) at (-1.5,0) {};
        \vertex [dot] (v) at (0,0) {};
        \vertex [] (b) at (1.5,0) {};
        \propag [glu] (a) to (v);
        \propag [glu] (v) to (b);
        \draw [scalar] (v) arc (-90:270:0.5) node[midway, above] {$\varphi^{(l,m)}$};
    \end{feynhand}
    \end{tikzpicture}
    +
    \begin{tikzpicture}[baseline={(current bounding box.center)}]
    \begin{feynhand}
        \vertex [] (a) at (-1.5,0) {};
        \vertex [] (b) at (1.5,0) {};
        \vertex [dot] (v1) at (-0.5,0) {};
        \vertex [dot] (v2) at (0.5,0) {};
        \propag [glu] (a) to (v1);
        \propag [glu] (v2) to (b);
        \draw [scalar] (v1) arc (180:0:0.5) node[midway, above] {$\varphi^{(l,m)}$};
        \draw [scalar] (v2) arc (0:-180:0.5) node[midway, below] {$\varphi^{(l,n)}$};
    \end{feynhand}
    \end{tikzpicture}
    = -\frac{1}{96\pi m_l^2} \Tr[T_I^{(l)} T_J^{(l)}] 
    \int_{\Sigma} d^2 x  \sqrt{\det g} H_{ab}^I H^{Jab},
\end{align}
where we omitted the higher-derivative terms and restored $\sqrt{\det g}$ to obtain the covariant result.
Using $\Tr[T_I^{(l)} T_J^{(l)}]= -l(l+1) (2l+1) \delta_{IJ}/3$, 
the covariant-derivative contribution to $c_{\text{SO(3)}}$ from $l \ge 2$ modes scales as
\begin{align}
    c_{\text{SO(3)}}^{l} \sim  (2l+1) r_0^2.
    \label{eq:l>1-modes-contribution}
\end{align}
The covariant-derivative contribution therefore does not vanish as $l \to \infty$,
but instead grows with $l$.
The $1/m_l^2 \sim l^{-2}$ suppression is offset by the multiplicity $2l+1$
and the quadratic Casimir $l(l+1)$, giving an $O(l)$ one-loop contribution.
Consequently, within this estimate, higher KK modes are not parametrically
suppressed at one loop, and the sum over the full KK tower is UV sensitive.
We emphasize, however, that Eq.~\eqref{eq:l>1-modes-contribution} neglects
mode-mixing contributions, which may modify the large-$l$ behavior;
a more complete analysis is left for future work.

We finally comment on two subtleties: boundary conditions and the Weyl anomaly.

\paragraph{Boundary conditions.}
A complete evaluation of the path integral~\eqref{eq:path-integral-massive-KK-modes}
requires specifying boundary conditions, including the quantities held fixed at the boundary
and the corresponding thermodynamic ensemble, together with appropriate boundary terms
in the action.
Since the spectrum of the covariant Laplacian $-D^2$ depends on the boundary conditions,
so do the propagator $(-D^2+m_1^2)^{-1}$ and the full tree-level contribution in
Eq.~\eqref{eq:tree-level-correction-from-l=1w/propagator}.
In the present analysis, however, we use the inverse-mass expansion of the propagator
and retain only the leading local term $1/m_1^2$, which does not capture the
boundary-dependent spectral information.
Likewise, at one loop, we retain only local bulk terms and do not include boundary
contributions.
Accordingly, the tree-level and one-loop terms evaluated here describe the local bulk
contributions, while boundary-condition dependence lies beyond the terms retained in
our approximation.
At the level of the local derivative expansion of the functional determinant,
dependence on the boundary conditions appears only through boundary terms;
see, for example, Ref.~\cite{Vassilevich:2003xt}.

\paragraph{Weyl anomaly.}
The conformal anomaly associated with the Weyl transformation~\eqref{eq:Weyl-transformation}
also contributes at one loop to the massless sector effective action.
A complete one-loop treatment of the massive mode contributions therefore requires including the Weyl anomaly, which we leave for future work.
The tree-level result~\eqref{eq:tree-level-correction-from-l=1}, however, is unaffected. Thus, our conclusion that the tree-level correction to the graviphoton kinetic coefficient is not parametrically suppressed near extremality remains unchanged.

\section{Conclusions}
\label{Sec:Conclusion}

We studied the two-dimensional effective theory obtained by reducing the four-dimensional Einstein--Maxwell theory on an RN black hole background 
and found that massive electromagnetic KK modes cannot be neglected near extremality.
The reduction retains the massless dilaton, metric, graviphotons, and two-dimensional photon together with massive electromagnetic KK modes.
Integrating out the massive electromagnetic KK modes gives local bulk
corrections to the massless sector effective theory.
We kept only the leading local terms and focused on corrections to the two-dimensional SO(3) and U(1) gauge couplings.
Within our approximation, neglecting the massive modes requires the conditions in
Eq.~\eqref{eq:truncation-condition-from-l=1} on the horizon radius and black hole charge. The second inequality in Eq.~\eqref{eq:truncation-condition-from-l=1} fails near extremality, where the $l=1$ tree-level correction becomes parametrically comparable to the classical graviphoton kinetic term.

The present analysis does not include boundary-condition-dependent effects or the Weyl anomaly.
The Weyl anomaly affects the one-loop effective action but not the tree-level result.
In addition, several other effects are omitted and may modify the quantitative criterion for
neglecting massive KK modes.
Massive gravitational KK modes can contribute to both the electromagnetic and graviphoton
gauge couplings, including charge-dependent corrections through interactions with the background
electric field.
Higher-derivative terms can also introduce additional charge dependence through the background
curvature.
Including these effects is necessary to establish a precise criterion for neglecting massive KK modes.

Rotating charged black holes introduce further complications. Rotation generates a nonzero graviphoton background, allowing massive-mode exchange to induce rotation-dependent corrections to the electromagnetic coupling and kinetic mixing between the photon and graviphoton. Neglecting the massive modes may therefore require sufficiently small angular momentum. Rotation may also induce nonzero backgrounds for the massive modes, so the classical consistency of the truncation must be examined. We leave extensions to rotating black holes for future work.

\section*{Acknowledgments}
\noindent
This work is supported by JST Grant No. JPMJPF2221 (YA), by JSPS KAKENHI Grant No. JP22K03601 (TH), and by JST SPRING, Japan Grant No. JPMJSP2123 (YM).

\appendix

\section{Details of dimensional reduction}
\label{App:Details-of-dimensional-reduction}

This appendix gives the details of the dimensional reduction. 
We first define the real spherical harmonics and then reduce the Einstein–Maxwell action to obtain Eqs.~\eqref{eq:I_massless} and \eqref{eq:gauge-fixed-Maxwell-action-1}.
See also Ref.~\cite{Abe:2026sai} for details of the dimensional reduction without graviphotons.

\subsection{Spherical harmonics}
\label{App:Spherical-harmonics}

We use real-valued spherical harmonics $Y^{\mathbb{R}}_{(l,m)}$, defined in terms of the standard complex spherical harmonics $Y^{\mathbb{C}}_{(l,m)}$ by
\begin{align}
    Y_{(l,m)}^{\mathbb{R}} (y) :=
    \begin{dcases}
        \frac{1}{\sqrt{2}}\left[Y^{\mathbb{C}}_{(l,-m)} (y) + (-1)^m Y^{\mathbb{C}}_{(l,m)} (y) \right] = \sqrt{2} \Re Y^{\mathbb{C}}_{(l,-m)}(y)& (m>0)
        \\
        Y^{\mathbb{C}}_{(l,0)} (y) & (m=0)
        \\
        \frac{i}{\sqrt{2}}\left[- (-1)^m Y^{\mathbb{C}}_{(l,-m)} (y) + Y^{\mathbb{C}}_{(l,m)} (y) \right] = -\sqrt{2} \Im Y^{\mathbb{C}}_{(l,m)}(y)& (m<0)
    \end{dcases}.
    \label{eq:def-of-real-valued-SH}
\end{align}

Since $S^2$ admits no harmonic one-forms, every smooth one-form on $S^2$ can be expanded in the following vector spherical harmonics with $l\geq1$:
\begin{align}
    \partial_\alpha Y^{\mathbb{R}}_{(l,m)},
    \qquad
    \epsilon_{\alpha\beta}(y)\partial^\beta Y^{\mathbb{R}}_{(l,m)}(y).
\end{align}
These vector spherical harmonics satisfy the eigenvalue equations
\begin{align}
    \Delta^{(1)}_{S^2}\partial_\alpha Y^{\mathbb{R}}_{(l,m)}
    =
    l(l+1)\partial_\alpha Y^{\mathbb{R}}_{(l,m)},
    \
    \Delta^{(1)}_{S^2}\epsilon_{\alpha\beta}\partial^\beta Y^{\mathbb{R}}_{(l,m)}
    =
    l(l+1)\epsilon_{\alpha\beta}\partial^\beta Y^{\mathbb{R}}_{(l,m)}.
    \label{eq:def-of-DY&EDY}
\end{align}
Here, the operator $\Delta^{(1)}_{S^2}$ acts on a one-form $v_\alpha$ as
\begin{align}
    \Delta^{(1)}_{S^2}v_\alpha
    := - \left( \nabla^2 h_{\alpha\beta} + [\nabla_\alpha,\nabla_\beta] \right) v^\beta.
\end{align}
The orthonormality condition~\eqref{eq:orthogonal-condition-SH} gives
\begin{align}
    \int_{S^2}d^2y\sqrt{\det h}~\partial_\alpha Y^{\mathbb{R}}_{(l,m)}\partial^\alpha Y^{\mathbb{R}}_{(l’,m’)}
    &=
    l(l+1)\delta_{ll’}\delta_{mm’},
    \label{eq:orthogonal-condition-DY}
    \\
    \int_{S^2}d^2y\sqrt{\det h}~
    \epsilon_{\alpha\beta}\partial^\beta Y^{\mathbb{R}}_{(l,m)}
    \epsilon^{\alpha\gamma}\partial_\gamma Y^{\mathbb{R}}_{(l’,m’)}
    &=
    l(l+1)\delta_{ll’}\delta_{mm’},
    \label{eq:orthogonal-condition-EDY&EDY}
    \\
    \int_{S^2}d^2y\sqrt{\det h}~
    \partial_\alpha Y^{\mathbb{R}}_{(l,m)}
    \epsilon^{\alpha\gamma}\partial_\gamma Y^{\mathbb{R}}_{(l’,m’)} &= 0.
    \label{eq:orthogonal-condition-DY&EDY}
\end{align}

\subsection{Einstein gravity}
\label{app:Einstein-Hilbert-action}
We next reduce the Einstein–Hilbert action and the Gibbons–Hawking–York boundary term. 
Like the electromagnetic fluctuations, the metric fluctuations can be expanded in spherical harmonics (see Ref.~\cite{Michelson:1999kn}):
\begin{align}
    \delta g_{ab}(x,y) &= \sum_{l=0}^\infty \sum_{|m|\le l} g_{ab}^{(l,m)}(x) Y^{\mathbb{R}}_{(l,m)}(y),
    \\
    \delta g_{a\alpha}(x,y) &= 
    \sum_{l=1}^\infty \sum_{|m|\le l} \frac{1}{m_l}
    \left[
        \tilde{B}_{a}^{(l,m)}(x) \partial_\alpha Y^{\mathbb{R}}_{(l,m)}(y)
        +
        B_{a}^{(l,m)}(x) \epsilon_{\alpha\beta} \partial^\beta Y^{\mathbb{R}}_{(l,m)}(y)
    \right],
    \\
    \delta g_{\alpha\beta}(x,y) &= 
    \sum_{l=2}^\infty \sum_{|m|\le l}
    \frac{1}{m_l^2}
    \left[
        \tilde{\Psi}^{(l,m)} (x)\nabla_{\{\alpha} \nabla_{\beta\}}  Y^{\mathbb{R}}_{(l,m)}(y)
        +
        \Psi^{(l,m)} (x) \nabla_{\{\alpha} \epsilon_{\beta\}\gamma} \partial^\gamma Y^{\mathbb{R}}_{(l,m)}(y)
     \right]
     \notag\\
     &\quad+\sum_{l=0}^\infty \sum_{|m|\le l} \Phi^{(l,m)}(x)Y^{\mathbb{R}}_{(l,m)}(y)h_{\alpha\beta}(y),
\end{align}
where the curly brackets $\{\}$ denote the symmetric traceless combination, $V_{\{\alpha}W_{\beta\}}:=V_\alpha W_\beta +V_\beta W_\alpha - h_{\alpha\beta} V_\gamma W^\gamma$ in two-dimensional space. 
Here, $\widetilde B_a^{(l,m)}$ and $B_a^{(l,m)}$ are the two-dimensional vector coefficients of the polar and axial vector harmonics, respectively,
while $\widetilde\Psi^{(l,m)}$ and $\Psi^{(l,m)}$ are two-dimensional scalar coefficients multiplying the corresponding traceless tensor harmonics.
The field $\Phi^{(l,m)}$ is the scalar coefficient of the trace mode.
The two-dimensional metric and dilaton fluctuations arise from the $l = 0$ modes $g_{ab}^{(0,0)}$ and $\Phi^{(0,0)}$, respectively.
The Killing vectors introduced in Eq.~(2.15) are the $l=1$ axial vector spherical harmonics,
\begin{align}
    \zeta_I^\alpha =\sqrt{\frac{4\pi}{3}} \epsilon^{\alpha\beta} \partial_\beta Y^{\mathbb{R}}_{(1,I)},
    \label{eq:VSH-Killing}
\end{align}
and hence the graviphotons $B_a^I$ correspond to the $l=1$ vector modes $B_a^{(1,I)}$. 
Their mass terms are forbidden by the local SO(3) gauge symmetry. 
We set all other massive gravitational KK fields to zero.  
After the Weyl transformation~\eqref{eq:Weyl-transformation}, the metric ansatz becomes
\begin{align}
    ds^2 = \frac{\Phi}{r_0^2} \left[ g_{ab}dx^a dx^b + r_0^2 h_{\alpha\beta} e^\alpha e^\beta \right].
    \label{eq:4D-metric-ansatz}
\end{align}

We perform the Kaluza--Klein reduction of the Einstein--Hilbert action~\eqref{eq:Einstein-Hilbert-action}.
For the metric ansatz~\eqref{eq:4D-metric-ansatz}, the four-dimensional scalar curvature is
\begin{align}
    R_{(4)} &= \frac{r_0^2}{\Phi}
    \left[
        R_{(2)} +\frac{2}{r_0^2} -\frac{r_0^2}{4}h_{\alpha\beta}\zeta^\alpha_I \zeta^\beta_J H^{I}_{ab} H^{Jab}
        +\frac{3}{2}
        \left(
            \frac{g^{ab}(\partial_a \Phi)(\partial_b \Phi)}{\Phi^2}
            -2\frac{g^{ab}\nabla_a \nabla_b \Phi}{\Phi}
        \right)
    \right],
    \label{eq:4D-scalar-curvature}
\end{align}
where $2/r_0^2$ is the scalar curvature of the sphere in the rescaled metric. 
From this formula, the Einstein–Hilbert action~\eqref{eq:Einstein-Hilbert-action} becomes
\begin{align}
     I_{\mr{EH}}
     &= - \frac{1}{4 G_N} \int_{\Sigma} d^2x \sqrt{\det g} \biggl[
        \Phi \Bigl(
            R_{(2)} + \frac{2}{r_0^2}
        \Bigr) + \frac{3}{2 \Phi} (\partial\Phi)^2 - 3\nabla^2 \Phi
    \biggr]
    \notag\\
    &\qquad + \frac{r_0^2}{24G_N}\int_{\Sigma} d^2x \sqrt{\det g}~\Phi H^I_{ab}H^{Iab},
    \label{eq:2D-Einstein-Hilbert-action}
\end{align}
where $R_{(2)}$ is the two-dimensional scalar curvature.
In deriving this expression, we used
\begin{align}
    \int_{S^2} d\Omega~ h_{\alpha\beta} \zeta^\alpha_I \zeta^\beta_J
    = \frac{8\pi}{3} \delta_{IJ},
    \label{eq:Integration-of-Killing-vectors}
\end{align}
which follows from Eqs.~\eqref{eq:orthogonal-condition-EDY&EDY} and \eqref{eq:VSH-Killing}.

We next reduce the Gibbons–Hawking–York term~\eqref{eq:Gibbons-Hawking-York-term}. 
The metric ansatz~\eqref{eq:4D-metric-ansatz} and the normalization of the unit normal vectors are related by
\begin{align}
	n^a_{(4)} = \frac{r_0}{\Phi^{1/2}} n^a_{(2)}.
\end{align}
From the definition of $K_{(3)}$, we find
\begin{align}
	K_{(3)} = \frac{r_0}{\Phi^{1/2}} \biggl[
		\frac{3}{2\Phi} n^a_{(2)} \partial_a \Phi + K_{(1)}
	\biggr],
\end{align}
where $K_{(1)} \coloneqq \nabla_a n^a_{(2)}$ is the extrinsic curvature of $\del \Sigma$.
The Gibbons--Hawking--York term \eqref{eq:Gibbons-Hawking-York-term} becomes
\begin{align}
	I_\GHY = - \frac{1}{2 G_N} \int_{\del \Sigma} dx \sqrt{\det \gamma} \Bigl(
		\Phi K_{(1)} + \frac{3}{2} n^a_{(2)} \del_a \Phi
	\Bigr).
	\label{eq:2D-Gibbons-Hawking-York-term}
\end{align}
The second term in Eq.~\eqref{eq:2D-Gibbons-Hawking-York-term} cancels the total-derivative term in Eq.~\eqref{eq:2D-Einstein-Hilbert-action}. 
The complete action for the gravitational sector is
\begin{align}
     I_{\mr{EH}} + I_\GHY
     &= - \frac{1}{4 G_N} \int_{\Sigma} d^2x \sqrt{\det g} \biggl[
        \Phi \Bigl(
            R_{(2)} + \frac{2}{r_0^2}
        \Bigr) + \frac{3}{2 \Phi} (\nabla\Phi)^2
    \biggr]
    - \frac{1}{2 G_N} \int_{\del \Sigma} dx \sqrt{\det \gamma} \Phi K_{(1)}
    \notag\\
    &\qquad + \frac{r_0^2}{24G_N}\int_{\Sigma} d^2x \sqrt{\det g}~\Phi H^I_{ab}H^{Iab}.
    \label{eq:2D-gravity-action}
\end{align}

\subsection{Maxwell action}
\label{App:Maxwell-action}

We now reduce the Maxwell action~\eqref{eq:Maxwell-action}. 
The fluctuating part of the four-dimensional field strength is defined by
\begin{align}
    \frac{1}{2}\delta F_{\mu\nu} dx^\mu \wedge dx^\nu &:= d ( \delta A_\mu dx^\mu),
\end{align}
where $d$ denotes the exterior derivative on $\mathcal{M}$.
Using Eq.~\eqref{eq:A_a&A_alpha} and 
\begin{align}
    de^\alpha &= \frac{1}{2}\zeta^\alpha_IH_{ab}^I dx^a \wedge dx^b 
    - (\partial_{\beta} \zeta^\alpha_I)B_a^I  dx^a \wedge e^{\beta},
    \label{eq:de}
\end{align}
we obtain the following field-strength components in the basis $(dx^a, e^\alpha)$:
\begin{align}
    \delta F_{\mu\nu} dx^\mu \wedge dx^\nu
    &= \delta F_{ab} dx^a \wedge dx^b + \delta F_{a\alpha} dx^a \wedge e^\alpha + \delta F_{\alpha a} e^\alpha \wedge dx^a + \delta F_{\alpha\beta} e^\alpha \wedge e^{\beta},
    \label{eq:F}
    \\
    \delta F_{ab} &=
    \left( \partial_a - B^I_a\zeta^\alpha_I\partial_\alpha \right) \delta A_b - \left( \partial_b - B^I_b\zeta^\alpha_I\partial_\alpha \right) \delta A_a
    +\zeta^\alpha_I \delta A_\alpha H^I_{ab},
    \\
    \delta F_{a\alpha}
    &= -\delta  F_{\alpha a} =\left(\partial_a - B^I_a\zeta^{\beta}_I\partial_{\beta}\right) \delta A_\alpha
    -B^I_a (\partial_\alpha \zeta^{\beta}_I) \delta A_{\beta} 
    -\partial_\alpha \delta A_a,
    \\
    \delta F_{\alpha\beta} &= \partial_{\alpha} \delta A_{\beta} - \partial_{\beta} \delta A_\alpha.
\end{align}
Substituting Eqs.~\eqref{eq:4D-gauge-field-A_a-decomposition} and \eqref{eq:4D-gauge-field-A_k-decomposition} gives
\begin{align}
    \delta F_{ab} 
    &=  
    \left( \delta f_{ab} + \sum_{(l',m')} \frac{1}{m_l'} H^I_{ab} (S_I)_{(0,0)(l',m')} \varphi^{(l',m')} \right)Y^{\mathbb{R}}_{(0,0)}
    \notag\\
    &\quad+
    \sum_{(l,m)}\left(
        f_{ab}^{(l,m)} 
        +\frac{1}{m_l}H^I_{ab}\sum_{|n|\leq l}(T_I^{(l)})_{mn} \tilde{\varphi}^{(l,n)}
        +\sum_{(l',m')}\frac{1}{m_l'}H^I_{ab}(S_I)_{(l,m)(l',m')} \varphi^{(l',m')}
    \right) Y^{\mathbb{R}}_{(l,m)},
    \label{eq:F_ab}
    \\
    \delta F_{a\alpha}
    &= \sum_{(l,m)}\frac{1}{m_l} 
    \Big[
        (D_a \tilde{\varphi}^{(l,m)} - m_l a_a^{(l,m)}) \nabla_\alpha Y^{\mathbb{R}}_{(l,m)}
        +(D_a\varphi^{(l,m)} )\epsilon_{\alpha\beta}\nabla^\beta Y^{\mathbb{R}}_{(l,m)}
    \Big],
    \label{eq:F_aalpha}
    \\
    \delta F_{\alpha\beta} &= \sum_{(l,m)}\frac{l(l+1)}{m_l} \epsilon_{\alpha\beta}\varphi^{(l,m)}Y^{\mathbb{R}}_{(l,m)},
    \label{eq:F_alphabeta}
\end{align}
where the eigenvalue equation~\eqref{eq:eigenvalue-eq-SH} has been used,
with $l\geq1$ and $l'\geq1$. Hereafter, unless otherwise stated, $l$ and $l'$ are understood to satisfy $l\geq1$ and $l'\geq1$.
In the basis $dx^a$ and $e^\alpha$, the metric~\eqref{eq:4D-metric-ansatz} is block diagonal, and the Maxwell action~\eqref{eq:Maxwell-action} becomes
\begin{align}
    I_{\text{Maxwell}}
    &= \frac{r_0^2}{4e^2} \int_{\Sigma} d^2x \sqrt{\det g} \int_{S^2} d^2y \sqrt{\det h}~
    \left[
        F_{ab} F^{ab} + \frac{2}{r_0^2}\delta F_{a\alpha} \delta F^{a\alpha} + \frac{1}{r_0^4}\delta F_{\alpha\beta} \delta F^{\alpha\beta}
    \right],
\end{align}
where the indices $a,b$ and $\alpha,\beta$ are raised using the two-dimensional metric $g^{ab}$ and $h^{\alpha\beta}$, respectively, and we write $F_{ab}:=F^{\text{RN}}_{ab} + \delta F_{ab}$ here.
Substituting Eqs.~\eqref{eq:F_ab}--\eqref{eq:F_alphabeta} into the above action and integrating over $S^2$, we obtain the reduced Maxwell action:
\begin{align}
    I_{\text{Maxwell}}
    & = \frac{r_0^2}{4e^2} \int_{\Sigma} d^2x \sqrt{\det g} \left(f_{ab} +\sum_{(l',m')}\frac{1}{m_l'} H^I_{ab}(S_I)_{(0,0)}{_{(l',m')}} \varphi^{(l',m')} \right)^2
    \notag\\
    &+
    \frac{r_0^2}{4e^2}\int_{\Sigma} d^2x \sqrt{\det g}
    \sum_{(l,m)}
    \Bigg[
    \left(
        f_{ab}^{(l,m)} 
        +\frac{1}{m_l}H^I_{ab}(T_I^{(l)})_m{_n} \tilde{\varphi}^{(l,n)}
        +\sum_{(l',m')}\frac{1}{m_l'}H^I_{ab}(S_I)_{(l,m)}{_{(l',m')}} \varphi^{(l',m')}
    \right)^2
    \notag\\
    &\quad
    +
        2(D_a \tilde{\varphi}^{(l,m)} - m_l a_a^{(l,m)})^2
        +2(D_a\varphi^{(l,m)} )^2
        +2m_l^2 (\varphi^{(l,m)})^2
    \Bigg],
    \label{eq:2D-reduced-Maxwell-action}
\end{align}
where we have used the orthogonality relations~\eqref{eq:orthogonal-condition-SH} and \eqref{eq:orthogonal-condition-DY}--\eqref{eq:orthogonal-condition-DY&EDY}.

The reduced Maxwell action~\eqref{eq:2D-reduced-Maxwell-action} has a local U(1) gauge symmetry implemented by transformations on the massive KK modes:
\begin{align}
    a_a \mapsto a_a + \partial_a \theta,
    \quad
    a_{a}^{(l,m)} \mapsto a_{a}^{(l,m)} + D_a \theta^{(l,m)},
    \quad
    \tilde{\varphi}^{(l,m)} \mapsto \tilde{\varphi}^{(l,m)} + m_l \theta^{(l,m)},
\end{align}
where $\theta$ and $\theta^{(l,m)}$ are the gauge transformation parameters.
These transformations originate from the four-dimensional U(1) gauge transformation $A_\mu \mapsto A_\mu + \partial_\mu \Theta$, and the four- and two-dimensional gauge transformation parameters are related as
\begin{align}
    \Theta(x,y)
    &= \theta(x) Y^{\mathbb{R}}_{(0,0)}
    + \sum_{(l,m)}
    \theta^{(l,m)}(x) Y^{\mathbb{R}}_{(l,m)}(y) ,
    \\
    d\Theta
    &= 
    (\partial_a\theta)Y^{\mathbb{R}}_{(0,0)}dx^a
    +
    \sum_{(l,m)}
    \left[
    D_a \theta^{(l,m)} Y^{\mathbb{R}}_{(l,m)} dx^a
    + \theta^{(l,m)} \partial_\alpha Y^{\mathbb{R}}_{(l,m)} e^\alpha
    \right] ,
\end{align}
where we used $\partial_\alpha Y_{(0,0)}=0$.
Under the U(1) transformation, $\varphi_{(l,m)}$ are invariant, whereas $f_{ab}^{(l,m)}$ transforms as 
\begin{align}
    f_{ab}^{(l,m)} \mapsto f_{ab}^{(l,m)} - H_{ab}^I \sum_{|n| \le l}(T_I^{(l)})_{mn} \theta^{(l,n)}.
\end{align}
These transformation laws make Eq.~\eqref{eq:2D-reduced-Maxwell-action} manifestly U(1) invariant.
To define the functional integral over the massive modes, we fix this gauge symmetry using the Faddeev--Popov procedure. We impose the gauge-fixing condition in Eq.~\eqref{eq:2D-gauge-fixing-condition} and introduce the Faddeev--Popov ghosts $(b^{(l,m)},c^{(l,m)})$, obtaining
\begin{align}
	I_{\mr{FP} + \mr{GF}} &=
	\sum_{(l,m)} \frac{r_0^2}{e^2} \int_\Sigma d^2x \sqrt{\det g}\biggl[
		b^{(l,m)} (- D_a D^a + m_l^2 ) c^{(l,m)} 
        + \frac{1}{2\xi^{(l,m)}} (
		  D_a a^{a(l,m)} - m_l \tilde{\varphi}^{(l,m)})^2
	\biggr],
    \label{Eq:FP+GF}
\end{align}
where $\xi^{(l,m)}$ are gauge fixing parameters.
For simplicity, we set $\xi^{(l,m)}=1$.
Combining Eqs.~\eqref{eq:2D-reduced-Maxwell-action} and \eqref{Eq:FP+GF} gives
\begin{align}
   & I_{\text{Maxwell}} + I_{\text{FP+GF}}
   \notag\\
    &= \frac{r_0^2}{4e^2} \int_{\Sigma} d^2x \sqrt{\det g}
    \left(f_{ab} +\sum_{(l',m')}\frac{1}{m_l'} H^I_{ab}(S_I)_{(0,0)(l',m')} \varphi^{(l',m')} \right)^2
    \notag\\
    &+
    \frac{r_0^2}{e^2}\int_{\Sigma} d^2x \sqrt{\det g}
    \sum_{(l,m)}
    \Bigg[
    b^{(l,m)}\Bigl(
			- D_a D^a + m_l^2
	\Bigr) c^{(l,m)} 
    \notag\\
    &\quad
    +\frac{1}{4}
    \left(
        f_{ab}^{(l,m)} 
        +\frac{1}{m_l} H^I_{ab} (T_I^{(l)})_{mn} \tilde{\varphi}^{(l,n)}
        +\sum_{(l',m')} \frac{1}{m_l'} H^I_{ab}(S_I)_{(l,m)(l',m')} \varphi^{(l',m')}
    \right)^2
    \notag\\
    &\quad
    + \frac{1}{2}
    \left(
        (D_a a^{a(l,m)})^2 + m_l^2 (a_a^{(l,m)})^2
        +(D_a\varphi^{(l,m)} )^2 + m_l^2 (\varphi^{(l,m)})^2
        +(D_a\tilde{\varphi}^{(l,m)} )^2 + m_l^2 (\tilde{\varphi}^{(l,m)})^2
    \right)
    \Bigg].
    \label{eq:gauge-fixed-Maxwell-action-2}
\end{align}
After integrating by parts, we obtain Eq.~\eqref{eq:gauge-fixed-Maxwell-action-1}.
Note that we have omitted the boundary terms in Eqs.~\eqref{eq:gauge-fixed-Maxwell-action-1} and \eqref{eq:gauge-fixed-Maxwell-action-2}.

\section{SO(3) gauge symmetry}
\label{App:SO(3)-gauge-symmetry}

In this appendix, we consider the SO(3) transformations of the KK modes and the corresponding representation matrices.

\subsection{SO(3) transformation}

The spherical harmonics transform under the SO(3) isometry~\eqref{eq:SO(3)-transformation-y} as follows:
\begin{align}
    Y_{(l,m)}^{\mathbb{R}}(y) \mapsto Y_{(l,m)}^{\mathbb{R}}(y) + \lambda^I (x) \zeta^\alpha_I (y) \partial_\alpha Y_{(l,m)}^{\mathbb{R}}(y)
    = Y_{(l,m)}^{\mathbb{R}}(y) + \lambda^I (x) \sum_{|n| \le l} (T_I^{(l)})_{nm} Y_{(l,n)}^{\mathbb{R}}(y),
    \label{eq:SO(3)-trans-SH}
\end{align}
where $T_I$ is defined by Eq.~\eqref{eq:def-of-T_I}.
The transformation mixes the magnetic quantum number $m$ but preserves $l$.
Indeed, $\zeta^\alpha_I \partial_\alpha Y_{(l,m)}^{\mathbb{R}}$ satisfies
\begin{align}
    \Delta^{(0)}_{S^2}\left(\zeta^\alpha_I \partial_\alpha Y_{(l,m)}^{\mathbb{R}}\right)
    =
    l(l+1)\left(\zeta^\alpha_I \partial_\alpha Y_{(l,m)}^{\mathbb{R}}\right).
\end{align}
Thus, the Killing-vector action preserves $l$ and mixes only $m$ within each fixed-$l$ sector. 
Using Eq.~\eqref{eq:SO(3)-trans-SH}, the expansion of $\delta A_a$ transforms as
\begin{align}
    \delta a_a Y_{(0,0)}^{\mathbb{R}} +  \sum_{(l,m)} a_{a}^{(l,m)} Y^{\mathbb{R}}_{(l,m)}
    &\mapsto
    \delta a_a Y_{(0,0)}^{\mathbb{R}} +  \sum_{(l,m)}  \left[ a_{a}^{(l,m)} + \sum_{|n| \le l}\lambda^I(T_I^{(l)})_{mn} a_{a}^{(l,n)} \right] Y^{\mathbb{R}}_{(l,m)}.
\end{align}
Therefore, the KK modes $a_a^{(l,m)}$ transform as
\begin{align}
    a_{a}^{(l,m)} \mapsto a_{a}^{(l,m)} + \lambda^I \sum_{|n| \le l}(T_I^{(l)})_{mn} a_{a}^{(l,n)}.
\end{align}
Similarly, the transformation of $\delta A_\alpha$ gives the scalar transformation laws:
\begin{align}
    \varphi^{(l,m)} \mapsto \varphi^{(l,m)} + \lambda^I \sum_{|n| \le l}(T_I^{(l)})_{mn} \varphi^{(l,n)},
    \quad
    \tilde{\varphi}^{(l,m)} \mapsto \tilde{\varphi}^{(l,m)} + \lambda^I \sum_{|n| \le l}(T_I^{(l)})_{mn} \tilde{\varphi}^{(l,n)}.
\end{align}

\subsection{Representation matrices}
\label{App:Representation-matrices}
The Killing vectors~\eqref{eq:def-of-Killing} are related to the standard angular momentum operators $L_{x,y,z}$ by
\begin{align}
    L_x = -i \zeta_{+1}, 
    \quad
    L_y = -i \zeta_{-1}, 
    \quad
    L_z = -i \zeta_{0}.
\end{align}
Since the ladder operators $L_{\pm}:=L_{x} \pm i L_{y}$ and $L_z$ act on the complex-valued spherical harmonics as
\begin{align}
    L_{\pm} Y^{\mathbb{C}}_{(l,m)} = \sqrt{(l \mp m) (l\pm m +1)} Y^{\mathbb{C}}_{(l,m \pm 1)},
    \quad
    L_{z} Y^{\mathbb{C}}_{(l,m)} =  mY^{\mathbb{C}}_{(l,m)},
\end{align}
we find
\begin{align}
    \zeta_{+1}^\alpha \partial_\alpha Y^{\mathbb{C}}_{(l,m)} &= \frac{i}{2}\left( \sqrt{(l-m)(l+m+1)}\delta_{n,m+1} + \sqrt{(l+m)(l-m+1)}\delta_{n,m-1} \right)Y^{\mathbb{C}}_{(l,n)},
    \\
    \zeta_{-1}^\alpha \partial_\alpha Y^{\mathbb{C}}_{(l,m)} &= \frac{1}{2}\left( \sqrt{(l-m)(l+m+1)}\delta_{n,m+1} - \sqrt{(l+m)(l-m+1)}\delta_{n,m-1} \right)Y^{\mathbb{C}}_{(l,n)},
    \\
    \zeta_{0}^\alpha \partial_\alpha Y^{\mathbb{C}}_{(l,m)} &= im \delta_{nm} Y^{\mathbb{C}}_{(l,n)},
\end{align}
where we have used $\zeta_{+1} = i(L_{+} + L_{-})/2$ and $\zeta_{-1} = (L_{+} - L_{-})/2$.
Combining the complex-harmonic transformation laws with Eq.~\eqref{eq:def-of-real-valued-SH}, we construct the real antisymmetric representation matrices $T_I$ defined in Eq.~\eqref{eq:def-of-T_I}.
In particular, for $l=1$, we find 
\begin{align}
    (T_I^{(1)})_{KJ} = -\epsilon_{IJK}.
\end{align}
The $l=1$ KK modes transform in the adjoint representation, and the generators satisfy
\begin{align}
    \tr[T_I^{(1)} T_J^{(1)}] = -2 \delta_{IJ}.
    \label{eq:T_I^1-trace}
\end{align}

\paragraph{Mixing matrix.}

We next evaluate the mixing matrix $S_I$, which is introduced by Eq.~\eqref{eq:def-of-S_I}.
Using $Y_{(0,0)}=1/\sqrt{4\pi}$, we obtain
\begin{align}
    (S_I)_{(0,0)}{_{(1,J)}}
    &= \int_{S^2} d\Omega Y^{\mathbb{R}}_{(0,0)} \zeta^\alpha_I \epsilon_{\alpha\beta} \nabla^\beta Y^{\mathbb{R}}_{(1,J)}
    = \frac{1}{\sqrt{3}}\int_{S^2} d\Omega (\epsilon^{\alpha\gamma} \nabla_\gamma Y^{\mathbb{R}}_{(1,I)}) (\epsilon_{\alpha\beta} \nabla^\beta Y^{\mathbb{R}}_{(1,J)}) = \sqrt{\frac{4}{3}} \delta_{IJ},
    \label{eq:S_I_00_1J}
\end{align}
where we have used Eqs.~\eqref{eq:def-of-S_I}, \eqref{eq:orthogonal-condition-EDY&EDY}, and \eqref{eq:VSH-Killing}.
In general, 
\begin{align}
    (S_I)_{(l',m')}{_{(l,m)}}
    &= \int_{S^2} d\Omega Y^{\mathbb{R}}_{(l',m')} \zeta^\alpha_I \epsilon_{\alpha\beta} \nabla^\beta Y^{\mathbb{R}}_{(l,m)}
    \notag\\
    &= \sqrt{\frac{4\pi}{3}} \int_{S^2} d\Omega Y^{\mathbb{R}}_{(l',m')} (\nabla_\alpha Y^{\mathbb{R}}_{(1,I)}) (\nabla^\alpha Y^{\mathbb{R}}_{(l,m)})
    \notag\\
    &=\sqrt{\frac{4\pi}{3}} \int_{S^2} d\Omega Y^{\mathbb{R}}_{(l',m')} 
    \frac{1}{2}\left[ \nabla^2 (Y^{\mathbb{R}}_{(1,I)}Y^{\mathbb{R}}_{(l,m)})- (\nabla^2 Y^{\mathbb{R}}_{(1,I)}) Y^{\mathbb{R}}_{(l,m)}-Y^{\mathbb{R}}_{(1,I)} (\nabla^2 Y^{\mathbb{R}}_{(l,m)})\right]
    \notag\\
    &=\sqrt{\frac{4\pi}{3}} \int_{S^2} d\Omega Y^{\mathbb{R}}_{(l',m')} 
    \frac{1}{2}\left[ \nabla^2 + 2 + l(l+1) \right] Y^{\mathbb{R}}_{(1,I)}Y^{\mathbb{R}}_{(l,m)}
    \notag\\
    &=\sqrt{\frac{4\pi}{3}} \sum_{l''=0}^\infty \sum_{|m''| \le l''} C^{\mathbb{R}}_{(1,I)(l,m)(l'',m'')} \int_{S^2} d\Omega Y^{\mathbb{R}}_{(l',m')} 
    \frac{1}{2}\left[ \nabla^2 + 2 + l(l+1) \right] Y^{\mathbb{R}}_{(l'',m'')}
    \notag\\
    &= \sqrt{\frac{4\pi}{3}} \sum_{l''=0}^\infty \sum_{|m''| \le l''} \frac{-l''(l''+1)+l(l+1)+2}{2}C^{\mathbb{R}}_{(1,I)(l,m)(l'',m'')} \int_{S^2} d\Omega Y^{\mathbb{R}}_{(l',m')} Y^{\mathbb{R}}_{(l'',m'')}
    \notag\\
    &=\sqrt{\frac{4\pi}{3}}\frac{-l'(l'+1)+l(l+1)+2}{2}C^{\mathbb{R}}_{(1,I)(l,m)(l',m')},
    \label{eq:S_I-CG}
\end{align}
where we used the eigenvalue equation~\eqref{eq:eigenvalue-eq-SH} and the orthogonality condition in Eq.~\eqref{eq:orthogonal-condition-SH}.
The Clebsch--Gordan coefficient $C^{\mathbb{R}}_{(1,I)(l,m)(l',m')}$ is defined by
\begin{align}
    \sum_{l'=0}^\infty \sum_{|m'| \le l'} C^{\mathbb{R}}_{(1,I)(l,m)(l',m')} Y_{(l',m')}^{\mathbb{R}} (y) := Y_{(1,I)}^{\mathbb{R}}(y) Y_{(l,m)}^{\mathbb{R}}(y).
\end{align}
The coefficients $C^{\mathbb{R}}_{(1,I)(l,m)(l',m')}$ vanish unless $l' = l\pm1$, so Eq.~\eqref{eq:S_I-CG} gives the same selection rule for $(S_I)_{(l',m')(l,m)}$.

\section{Evaluation of effective action}
\label{App:Evaluation-of-effective-action}

In this appendix, we rederive the tree-level and one-loop contributions in
Eqs.~\eqref{eq:tree-level-correction-from-l=1} and
\eqref{eq:1-loop-level-correction-from-l=1} by integrating out the $l=1$
Maxwell KK modes using Gaussian functional integrals, rather than evaluating
the corresponding Feynman diagrams directly as in Sec.~\ref{sec:Effective-action}.
This provides an alternative derivation of the same results.

\subsection{Functional integrals}

Retaining the massless sector and the $l=1$ Maxwell KK modes, while omitting interactions involving the $l \ge 2$ modes, we obtain
\begin{align}
    I&=I_{\text{massless}} + I_{\varphi} + I_{\tilde{\varphi}} + I_{a} + I_{\text{FP}}^{l=1},
    \\
    I_{\varphi} &:= \frac{r_0^2}{2e^2} \sum_{I,J=\pm1,0}\int_{\Sigma} d^2x \sqrt{\det g} \left[ \varphi^I \mathcal{K}_{IJ}\varphi^J  + \frac{1}{m_1} f^{ab} H_{ab}^I (S_I)_{(0,0)}{_{(1,J)}}\varphi^J \right],
    \label{eq:I_varphi}
    \\
    \mathcal{K}_{IJ} &:= \delta_{IJ} (-D^2 + m_1^2 ) + \frac{1}{2m_1^2}H^{K}_{ab} H^{Lab} \sum_{l=0}^2 \sum_{|m|\le l}(S_{K})_{(l,m)}{_{(1,I)}} (S_{L})_{(l,m)}{_{(1,J)}},
    \\
    I_{\tilde{\varphi}} &:= \frac{r_0^2}{2e^2} \sum_{I,J=\pm1,0} \int_{\Sigma} d^2x \sqrt{\det g} \left[ \tilde{\varphi}^I \tilde{\mathcal{K}}_{IJ} \tilde{\varphi}^J  + \frac{1}{m_1} \sum_{K=\pm1,0} f^{ab K} H_{ab}^I (T^{(1)}_I)_K{_J}\tilde{\varphi}^J \right],
    \label{eq:I_tilde-varphi}
    \\
    \tilde{\mathcal{K}}_{IJ} &:= \delta_{IJ} (-D^2 + m_1^2 ) + \frac{1}{2m_1^2}H^{I'}_{ab} H^{J'ab} \sum_{K=\pm1,0}(T^{(1)}_{I'})_K{_I}(T^{(1)}_{J'})_K{_J},
    \\
    I_{a} &:= \frac{r_0^2}{2e^2} \sum_{I,J=\pm1,0} \int_{\Sigma} d^2x \sqrt{\det g} ~a_a^I \left[  \delta_{IJ}g^{ab} \left(-D^2 + m_1^2 + \frac{1}{2}R_{(2)}\right) + H^{abK} (T_K^{(1)})_{IJ} \right] a_b^J,
    \label{eq:I_a}
    \\
    I_{\text{FP}}^{l=1} &:= \frac{r_0^2}{e^2} \sum_{I,J=\pm1,0} \int_{\Sigma} d^2x \sqrt{\det g} ~b^I \delta_{IJ} (- D^2 + m_1^2 ) c^J,
\end{align}
where $D^2:=D_a D^a$, and we used integration by parts and omitted boundary terms.
$\varphi^I$ is defined by
\begin{align}
    \varphi^I := \varphi^{(1,I)}, 
    \quad I=\pm1,0.
\end{align}
$\tilde{\varphi}^I$, $a_a^I$, $b^I$, and $c^I$ are defined similarly.
The action is quadratic in the massive fields. 
Performing the Gaussian integrals, including the mixing between $a_a^I$ and $\tilde{\varphi}^I$, gives
\begin{align}
    \int[d\varphi^I d\tilde{\varphi}^I da_a^I db^I dc^I]
    e^{- I_{\varphi} - I_{\tilde{\varphi}} - I_{a} - I_{\text{FP}}^{l=1}}
    = \frac{\det [-D^2 +m_1^2]}{\sqrt{\det [\mathcal{K}] \det [\tilde{\mathcal{K}}] \det [\hat{\mathcal{K}}]}} e^{-\Delta W_{\text{tree}}},
    \label{eq:massive-mode-partition-function}
\end{align}
where we defined
\begin{align}
    \hat{\mathcal{K}}_{IJ}^{ab} &:= \delta_{IJ} g^{ab} \left(-D^2 + m_1^2 +\frac{1}{2} R_{(2)}\right)  + H^{abK} (T_K^{(1)})_{IJ}
    \nn \\
    &+ \frac{1}{2m_1^2} g^{ab}D^e\left[ H_{cd}^{I'} (T_{I'}^{(1)})_{IK} (\tilde{\mathcal{K}}^{-1})_{KL} H^{cdJ'} (T_{J'}^{(1)})_{JL} \right] D_e 
    \notag\\
    &- \frac{1}{2m_1^2} D^b\left[ H_{cd}^{I'} (T_{I'}^{(1)})_{IK} (\tilde{\mathcal{K}}^{-1})_{KL} H^{cdJ'} (T_{J'}^{(1)})_{JL} \right] D^a,
    \label{eq:def-of-K-hat}
    \\
    \Delta W_{\text{tree}} &:= -\frac{r_0^2}{2e^2} \frac{1}{3m_1^2} \int_\Sigma d^2x \sqrt{\det g} ~f^{ab} H_{ab}^I (\mathcal{K}^{-1})_{IJ} f^{cd} H_{cd}^J.
    \label{eq:delta-W}
\end{align}
Equation~\eqref{eq:massive-mode-partition-function} results from the Gaussian
integration over the $l=1$ massive Maxwell KK fields $\varphi^I$,
$\widetilde{\varphi}^I$, and $a_a^I$, together with the associated FP ghosts
$b^I$ and $c^I$, while the massless fields are treated as backgrounds.
The determinant ratio in Eq.~\eqref{eq:massive-mode-partition-function}
gives the one-loop contribution.
Here, $\det$ denotes the determinant over the functional space and the
SO(3) representation indices $I,J$, and, where applicable, over the
spacetime vector indices $a,b$.
The effective vector operator $\widehat{\mathcal K}_{IJ}^{ab}$ in
Eq.~\eqref{eq:def-of-K-hat} is obtained by completing the square in the
coupled $(a_a^I,\widetilde{\varphi}^I)$ sector of
Eq.~\eqref{eq:I_tilde-varphi}.
The factor $e^{-\Delta W_{\rm tree}}$ in
Eq.~\eqref{eq:massive-mode-partition-function} arises from completing
the square for $\varphi^I$.
Using Eq.~\eqref{eq:S_I_00_1J}, this gives Eq.~\eqref{eq:delta-W}.
Here, $\mathcal K^{-1}$ denotes the inverse operator of $\mathcal K$,
satisfying
$\mathcal K_{IK}(\mathcal K^{-1})_{KJ}=\delta_{IJ}$.

Using the expansion
\begin{align}
    (\mathcal{K}^{-1})_{IJ} &\approx \frac{1}{m_1^2} \delta_{IJ} + \frac{1}{m_1^4}\delta_{IJ} D^2 - \frac{1}{2m_1^6} H^{I'}_{ab} H^{J'ab} \sum_{l=0}^2 \sum_{|m|\le l} (S_{I'})^{(l,m)}{_{(1,I)}} (S_{J'})^{(l,m)}{_{(1,J)}}+ \cdots,
\end{align}
the leading term in the inverse-mass expansion of $\Delta W_{\text{tree}}$ is
\begin{align}
    \Delta W_{\text{tree}} &\approx-\frac{r_0^2}{2e^2} \frac{1}{3m_1^4} \int_\Sigma d^2x \sqrt{\det g} ~ (f^{ab} H_{ab}^I)^2.
    \label{eq:Delta-W-tree}
\end{align}
Using the background electric flux in Eq.~\eqref{eq:f-decomposion}, together with
\begin{align}
    \left(\epsilon^{ab}H_{ab}^I\right)^2
    =2H_{ab}^IH_I^{ab},
    \qquad
    m_1^2=\frac{2}{r_0^2},
\end{align}
Eq.~\eqref{eq:Delta-W-tree} becomes
\begin{align}
    \Delta W_{\mathrm{tree}} \approx
    \frac{r_0^2e^2Q^2}{48\pi}
    \int_\Sigma d^2x\sqrt{\det g}\,
    H_{ab}^IH^{Iab}.
\end{align}
The resulting expression reproduces Eq.~\eqref{eq:tree-level-correction-from-l=1}. 
We omit terms cubic and higher order in the massless fluctuations and next evaluate the one-loop contribution.

\subsection{One-loop determinants}
\label{app:One-loop-determinants}

On a flat Euclidean spacetime background, we evaluate the contribution to $c_{\mathrm{SO}(3)}$ in Eq.~\eqref{eq:Delta_W} from the one-loop determinants in Eq.~\eqref{eq:massive-mode-partition-function}.
Throughout this subsection, spacetime indices are raised and lowered with the flat Euclidean metric $\delta_{ab}$.
\begin{align}
    Z_{\text{1-loop}}:=\frac{\det [-D^2 +m_1^2]}{\sqrt{\det [\mathcal{K}] \det [\tilde{\mathcal{K}}] \det [\hat{\mathcal{K}}]}}.
\end{align}
The terms in $-\log Z_{\text{1-loop}}$ that are quadratic in $B_a^I$
can be written in Fourier space as
\begin{align}
    -\log Z_{\text{1-loop}} 
    &= 
    c_{\mathrm{SO}(3)}^{\text{1-loop}} \int\frac{d^2k}{(2\pi)^2} H_{ab}^I (k) H^{Iab}(-k) + \cdots
    \\
    &=
    c_{\mathrm{SO}(3)}^{\text{1-loop}} \int\frac{d^2k}{(2\pi)^2}2B_a^I(k)B_b^I(-k)\left(k^2\delta^{ab}-k^ak^b\right) + \cdots,
\end{align}
where the ellipsis denotes higher-derivative terms.

We now evaluate the quadratic terms and determine $c_{\mathrm{SO}(3)}^{\text{1-loop}}$.
We first consider the massive fields $\varphi^I$, $\tilde{\varphi}^I$, $b^I$, and $c^I$.
The corresponding one-loop determinants in $-\log Z_{\text{1-loop}}$ generally take the following form:
\begin{align}
    \Gamma_{\text{scalar}}[n,C] &:= \frac{n}{2}\log \det \mathcal{O} = \frac{n}{2}\Tr\log \mathcal{O},
    \\
    \mathcal{O}_{IJ} &:=(-D^2 + m_1^2)\delta_{IJ} +  \frac{1}{2m_1^2}C_{IJKL} H^K_{ab} H^{Lab},
\end{align}
where $\det$ denotes the determinant taken over the functional space and the representation space.
For instance, the FP determinant corresponds to $n=-2$ and $C_{IJKL}=0$.
The $\varphi^I$ and $\tilde{\varphi}^I$ contributions correspond to $n=1$ and 
\begin{align}
    C_{IJKL} &= \sum_{l=0}^2 \sum_{|m|\le l}(S_{K})_{(l,m)}{_{(1,I)}} (S_{L})_{(l,m)}{_{(1,J)}},
    \\
    C_{IJKL} &= \sum_{K'=\pm1,0}(T^{(1)}_{K})_{K'}{_I}(T^{(1)}_{L})_{K'}{_J},
\end{align}
respectively.

Repeated indices such as $I,J$ are summed over $0,\pm1$.
Since 
\begin{align}
    D^2 \delta_{IJ}= \partial^2 \delta_{IJ} - (\partial^a B_a^{I'}) (T_{I'}^{(1)})_{IJ} - 2 B_a^{I'} (T_{I'}^{(1)})_{IJ} \partial^a + B^{I'}_a B^{J'a} (T_{I'}^{(1)})_{IK} (T_{J'}^{(1)})_{KJ},
\end{align} 
on the flat background, the operator can be written as
\begin{align}
    \mathcal{O}_{IJ} &:= (-\partial^2 + m_1^2) \delta_{IJ} + \mathcal{O}_{IJ}^{(1)} + \mathcal{O}_{IJ}^{(2)} + \frac{1}{2m_1^2}C_{IJKL} H^K_{ab} H^{Lab},
    \label{eq:O_IJ}
    \\
    \mathcal{O}_{IJ}^{(1)} &:=  (\partial^a B_a^{I'}) (T_{I'}^{(1)})_{IJ} + 2 B_a^{I'} (T_{I'}^{(1)})_{IJ} \partial_a,
    \\
    \mathcal{O}_{IJ}^{(2)} &:= -B^{I'}_a B^{J'a} (T_{I'}^{(1)})_{IK} (T_{J'}^{(1)})_{KJ}.
\end{align}
The second and third terms in Eq.~\eqref{eq:O_IJ} correspond to the vertices given in Eqs.~\eqref{eq:B_varphi_vaprhi}--\eqref{eq:B_b_c} and Eqs.~\eqref{eq:B_B_varphi_varphi}-\eqref{eq:B_B_b_c}, respectively.
The fourth term corresponds to the vertices given in Eqs.~\eqref{eq:mode-mixing-HH-varphi-varphi} and \eqref{eq:mode-mixing-HH-tildevarphi-tildevarphi}.
The one-loop determinant can be expanded as 
\begin{align}
    \Gamma_{\text{scalar}} &= 
    \frac{n}{2}\Tr \log [(-\partial^2 + m_1^2) \delta_{IJ}]
    + \frac{n}{2}\Tr \log [ \delta_{IJ} + (-\partial^2 + m_1^2)^{-1} (\mathcal{O}_{IJ}^{(1)} + \mathcal{O}_{IJ}^{(2)} + \frac{1}{2m_1^2}C_{IJKL} H^K_{ab} H^{Lab})]
    \notag\\
    &=\frac{n}{2}\Tr \log [(-\partial^2 + m_1^2) \delta_{IJ}]
    + \frac{n}{2}\Tr [ (-\partial^2 + m_1^2)^{-1} (\mathcal{O}_{IJ}^{(1)} + \mathcal{O}_{IJ}^{(2)} + \frac{1}{2m_1^2}C_{IJKL} H^K_{ab} H^{Lab})]
    \notag\\
    &- \frac{n}{4} \Tr [(-\partial^2 + m_1^2)^{-1} (\mathcal{O}_{IK}^{(1)} + \mathcal{O}_{IK}^{(2)} + \frac{1}{2m_1^2}C_{IKK''L} H^{K''}_{ab} H^{Lab})
    \notag\\
    &\times(-\partial^2 + m_1^2)^{-1} (\mathcal{O}_{KJ}^{(1)} + \mathcal{O}_{KJ}^{(2)}  + \frac{1}{2m_1^2}C_{KJK'L'} H^{K'}_{ab} H^{L'ab})] + \cdots,
    \label{eq:Gamma-scalar}
\end{align}
where $\Tr$ denotes the functional trace as well as the trace over the indices $I,J$, and the ellipsis denotes terms which start at cubic order in the background graviphoton and do not contribute to the quadratic calculation considered here.
The quadratic terms in $B_a$ in Eq.~\eqref{eq:Gamma-scalar}, which contribute to $c_{\text{SO(3)}}$, are 
\begin{align}
    \Gamma_{\text{scalar}}^{(1)} &:= -\frac{n}{4} \Tr [(-\partial^2 + m_1^2)^{-1} \mathcal{O}_{IK}^{(1)}(-\partial^2 + m_1^2)^{-1} \mathcal{O}_{KJ}^{(1)}],
    \\
    \Gamma_{\text{scalar}}^{(2)} &:= \frac{n}{2}\Tr [(-\partial^2 + m_1^2)^{-1} \mathcal{O}_{IJ}^{(2)}],
    \\
    \Gamma_{\text{scalar}}^{(3)} &:= \frac{n}{2}\Tr [(-\partial^2 + m_1^2)^{-1} \frac{1}{2m_1^2}C_{IJKL} H^{K}_{ab} H^{Lab}],
\end{align}

$\Gamma_{\text{scalar}}^{(1)}$ and $\Gamma_{\text{scalar}}^{(2)}$ correspond to a bubble diagram and a tadpole diagram, respectively:
\begin{align}
    \Gamma_{\text{scalar}}^{(1)} =
    \begin{tikzpicture}[baseline={(current bounding box.center)}]
    \begin{feynhand}
        \vertex [] (a) at (-1.5,0) {};
        \vertex [] (b) at (1.5,0) {};
        \vertex [dot] (v1) at (-0.5,0) {};
        \vertex [dot] (v2) at (0.5,0) {};
        \propag [glu] (a) to (v1);
        \propag [glu] (v2) to (b);
        \draw [scalar] (v1) arc (180:0:0.5);
        \draw [scalar] (v2) arc (0:-180:0.5);
    \end{feynhand}
    \end{tikzpicture}
    &=-\frac{n}{2} \int \frac{d^2 k}{(2\pi)^2} B^I_a(k) B^I_b(-k) \int \frac{d^2 p}{(2\pi)^2} \frac{2p^a + k^a}{p^2 + m_1^2} \frac{2p^b + k^b}{(p+k)^2 + m_1^2},
    \label{eq:Gamma-scalar-1-diagram}
    \\
    \Gamma_{\text{scalar}}^{(2)} =
    \begin{tikzpicture}[baseline={(current bounding box.center)}]
    \begin{feynhand}
        \vertex [](a) at (-1.5,0) {};
        \vertex [dot] (v) at (0,0) {};
        \vertex [] (b) at (1.5,0) {};
        \propag [glu] (a) to (v);
        \propag [glu] (v) to (b);
        \draw [scalar] (v) arc (-90:270:0.5);
    \end{feynhand}
    \end{tikzpicture}
    &= n \int \frac{d^2 k}{(2\pi)^2} B^I_a(k) B^{Ia}(-k) \int \frac{d^2 p}{(2\pi)^2} \frac{1}{p^2 + m_1^2},
    \label{eq:Gamma-scalar-2-diagram}
\end{align}
where we used Eq.~\eqref{eq:T_I^1-trace}.
For simplicity, dashed lines below denote propagators of both real scalars and ghosts.
The sum $\Gamma_{\text{scalar}}^{(1)} + \Gamma_{\text{scalar}}^{(2)}$ is finite in two dimensions and has the low-momentum expansion:
\begin{align}
    \Gamma_{\text{scalar}}^{(1)} + \Gamma_{\text{scalar}}^{(2)} 
    &= \frac{n}{48\pi m_1^2}\int \frac{d^2 k}{(2\pi)^2} 2B_a^I(k) B_b^I(-k) (k^2 \delta^{ab} -k^a k^b )(1+O(k^2/m_1^2)).
    \label{eq:Gamma^1+Gamma^2-scalar}
\end{align}

$\Gamma_{\text{scalar}}^{(3)}$ corresponds to the following tadpole diagram:
\begin{align}
    \Gamma_{\text{scalar}}^{(3)} =
    \begin{tikzpicture}[baseline={(current bounding box.center)}]
    \begin{feynhand}
        \vertex [](a) at (-1.5,0) {};
        \vertex [ringdot] (v) at (0,0) {};
        \vertex [] (b) at (1.5,0) {};
        \propag [glu] (a) to (v);
        \propag [glu] (v) to (b);
        \draw [scalar] (v) arc (-90:270:0.5);
    \end{feynhand}
    \end{tikzpicture}
    &= \frac{n}{4m_1^2} \sum_{I'=\pm1,0}C_{I'I'IJ} \int \frac{d^2 k}{(2\pi)^2} H^I_{ab}(k) H^{Jab}(-k) \int \frac{d^2 p}{(2\pi)^2} \frac{1}{p^2 + m_1^2}.
    \label{eq:Gamma-scalar-3}
\end{align}
In two dimensions, the $p$-integral diverges, so we use zeta-function regularization.
We first introduce a regulator $\epsilon$ and
\begin{align}
    I (\epsilon) := \mu^{2\epsilon}\int \frac{d^2 p}{(2\pi)^2} \frac{1}{(p^2 + m_1^2)^{1+\epsilon}},
\end{align}
where the mass parameter $\mu$ keeps the integral dimensionless.
The original $p$-integral in Eq.~\eqref{eq:Gamma-scalar-3} is reproduced in the $\epsilon\to0$ limit.
Introducing the Schwinger-time $s$, $I(\epsilon)$ can be written as
\begin{align}
    I (\epsilon) = \frac{\mu^{2\epsilon}}{\Gamma(1+\epsilon)} \int \frac{d^2 p}{(2\pi)^2}\int_0^\infty ds s^{\epsilon}e^{-s(p^2+m_1^2)}.
\end{align}
Performing the $p$ integral followed by the $s$ integral gives
\begin{align}
    I (\epsilon) &= \frac{\mu^{2\epsilon}}{4\pi\Gamma(1+\epsilon)} \int_0^\infty ds s^{\epsilon-1} e^{-sm_1^2}
    =
    \frac{1}{4\pi}
    \left(\frac{\mu^2}{m_1^2}\right)^\epsilon
    \frac{\Gamma\left(\epsilon\right)}{\Gamma(1+\epsilon)}
    \notag\\
    &=
    \frac{1}{4\pi}
    \left[
    \frac{1}{\epsilon}
    +\log\frac{\mu^2}{m_1^2}
    \right]
    +O(\epsilon).
\end{align}
The pole $1/\epsilon$ is absorbed into an appropriate local counterterm.
After subtracting the pole and taking $\epsilon\to0$,
\begin{align}
    I_{\text{ren.}} = \frac{1}{4\pi} \log\frac{\mu^2}{m_1^2}.
    \label{eq:I_ren}
\end{align}
We choose $\mu=m_1$.
In this subtraction scheme, $\Gamma_{\text{scalar}}^{(3)}$ vanishes,
so the contribution to the graviphoton kinetic term is independent of
$C_{IJKL}$.
\begin{align}
    \Gamma_{\text{scalar}} \approx \frac{n}{48\pi m_1^2} \int_{\Sigma} d^2x H_{ab}^I H^{Iab}.
\end{align}
The $\varphi^I$ and $\tilde{\varphi}^I$ contributions  correspond to $n=1$, while the ghost contributions  from $c^I$ and $b^I$ correspond to $n=-2$.
The scalar and ghost contributions to the quadratic graviphoton kinetic term cancel for $\mu=m_1$. 
The remaining one-loop contribution therefore comes from the determinant of
the effective vector operator $\widehat K_{IJ}^{ab}$.

We next evaluate the massive-vector contribution from $a^I_a$:
\begin{align}
    \Gamma_{\text{vector}} &:= \frac{1}{2} \log \det \hat{\mathcal{K}}_{IJ}^{ab} = \frac{1}{2}\Tr \log \hat{\mathcal{K}}_{IJ}^{ab}.
    \label{eq:Gamma-vactor}
\end{align}
Here $\Tr$ denotes the functional trace together with the traces over the representation indices and the spacetime vector indices, with the latter
contracted using the spacetime metric $\delta_{ab}$.
The operator $\hat{\mathcal{K}}$ can be expanded as
\begin{align}
    \hat{\mathcal{K}}_{IJ}^{ab} &= \delta_{IJ} \delta^{ab} (- \partial^2 + m_1^2) + \delta^{ab}  (\mathcal{O}^{(1)}_{IJ} + \mathcal{O}^{(2)}_{IJ} ) + H^{abK} (T_K^{(1)})_{IJ}
    \notag\\
    &+ \frac{1}{2m_1^2} H_{cd}^{I'} (T_{I'}^{(1)})_{IK} (\tilde{\mathcal{K}}^{-1})^{KL} H^{cdJ'} (T_{J'}^{(1)})_{JL}
    ( \partial^2 \delta^{ab} - \partial^a \partial^b) + \cdots,
    \label{eq:expansion-of-K-hat}
\end{align}
where the ellipsis denotes terms of higher order in $B_a^I$ and higher-derivative terms that do
not contribute to the leading quadratic gauge kinetic term considered here.
Since the last term in Eq.~\eqref{eq:expansion-of-K-hat} is already quadratic in the background graviphoton, we use the zeroth-order approximation
$(\widetilde{\mathcal K}^{-1})_{IJ} =\delta_{IJ}(-\partial^2+m_1^2)^{-1}$.
The quadratic terms in $B_a^I$ contributing to $c_{\text{SO(3)}}$ are
\begin{align}
    \Gamma_{\text{vector}}^{(1)} &:= -\frac{1}{4} \Tr [\delta^{ab}(-\partial^2 + m_1^2)^{-1} \mathcal{O}_{IK}^{(1)}(-\partial^2 + m_1^2)^{-1} \mathcal{O}_{KJ}^{(1)}],
    \\
    \Gamma_{\text{vector}}^{(2)} &:= \frac{1}{2}\Tr [(-\partial^2 + m_1^2)^{-1} \delta^{ab}\mathcal{O}_{IJ}^{(2)}],
    \\
    \Gamma_{\text{vector}}^{(3)} &:= -\frac{1}{4}\Tr [(-\partial^2 + m_1^2)^{-1} H_{ac}^I T_I^{(1)}(-\partial^2 + m_1^2)^{-1} H{^J}{^c}{_b} T_J^{(1)}],
    \\
    \Gamma_{\text{vector}}^{(4)} &:= \frac{1}{4m_1^2}\Tr  [ (-\partial^2 + m_1^2)^{-1} H_{cd}^{I'} (T_{I'}^{(1)})_{IK} (-\partial^2 + m_1^2)^{-1} H^{J'cd} (T_{J'}^{(1)})_{JK}
    ( \partial^2 \delta^{ab} - \partial^a \partial^b)].
\end{align}
$\Gamma_{\text{vector}}^{(1)}$ and $\Gamma_{\text{vector}}^{(2)}$ correspond to the following diagrams:
\begin{align}
    \Gamma_{\text{vector}}^{(1)} =
    \begin{tikzpicture}[baseline={(current bounding box.center)}]
    \begin{feynhand}
        \vertex [] (a) at (-1.5,0) {};
        \vertex [] (b) at (1.5,0) {};
        \vertex [dot] (v1) at (-0.5,0) {};
        \vertex [dot] (v2) at (0.5,0) {};
        \propag [glu] (a) to (v1);
        \propag [glu] (v2) to (b);
        \draw [boson] (v1) arc (180:0:0.5) node[midway, above] {$a_a^I$};
        \draw [boson] (v2) arc (0:-180:0.5) node[midway, below] {$a_b^J$};
    \end{feynhand}
    \end{tikzpicture}
    &=-\int \frac{d^2 k}{(2\pi)^2} B^I_a(k) B^I_b(-k) \int \frac{d^2 p}{(2\pi)^2} \frac{2p^a + k^a}{p^2 + m_1^2} \frac{2p^b + k^b}{(p+k)^2 + m_1^2},
    \\
    \Gamma_{\text{vector}}^{(2)} =
    \begin{tikzpicture}[baseline={(current bounding box.center)}]
    \begin{feynhand}
        \vertex [](a) at (-1.5,0) {};
        \vertex [dot] (v) at (0,0) {};
        \vertex [] (b) at (1.5,0) {};
        \propag [glu] (a) to (v);
        \propag [glu] (v) to (b);
        \draw [boson] (v) arc (-90:270:0.5) node[midway, above] {$a_a^I$};
    \end{feynhand}
    \end{tikzpicture}
    &= 2\int \frac{d^2 k}{(2\pi)^2} B^I_a(k) B^{Ia}(-k) \int \frac{d^2 p}{(2\pi)^2} \frac{1}{p^2 + m_1^2},
\end{align}
where we used Eq.~\eqref{eq:T_I^1-trace}.
In two dimensions, $\Gamma_{\text{vector}}^{(1)}+\Gamma_{\text{vector}}^{(2)}$
is finite.
By gauge invariance, the sum of these contributions must be transverse.
Their UV divergences cancel, and expanding for $k^2\ll m_1^2$, we obtain
\begin{align}
    \Gamma_{\text{vector}}^{(1)} + \Gamma_{\text{vector}}^{(2)} 
    &= \frac{1}{24\pi m_1^2}\int \frac{d^2 k}{(2\pi)^2} 2B_a^I(k) B_b^I(-k) (k^2 \delta^{ab} -k^a k^b )(1+O(k^2/m_1^2)).
    \label{eq:Gamma^1+Gamma^2-vector}
\end{align}

$\Gamma_{\text{vector}}^{(3)}$ corresponds to the following diagram:
\begin{align}
    \Gamma_{\text{vector}}^{(3)} =
    \begin{tikzpicture}[baseline={(current bounding box.center)}]
    \begin{feynhand}
        \vertex [] (a) at (-1.5,0) {};
        \vertex [] (b) at (1.5,0) {};
        \vertex [ringdot] (v1) at (-0.5,0) {};
        \vertex [ringdot] (v2) at (0.5,0) {};
        \propag [glu] (a) to (v1);
        \propag [glu] (v2) to (b);
        \draw [boson] (v1) arc (180:0:0.5) node[midway, above] {$a_a^I$};
        \draw [boson] (v2) arc (0:-180:0.5) node[midway, below] {$a_b^J$};
    \end{feynhand}
    \end{tikzpicture}
    &= -\frac{1}{2} \int \frac{d^2 k}{(2\pi)^2} H^I_{ab}(k) H^{Iab}(-k) \int \frac{d^2 p}{(2\pi)^2} \frac{1}{p^2 + m_1^2} \frac{1}{(p+k)^2 + m_1^2}
\end{align}
The $p$-integral is
\begin{align}
    \int \frac{d^2 p}{(2\pi)^2} \frac{1}{p^2 + m_1^2} \frac{1}{(p+k)^2 + m_1^2}
    = \frac{1}{4\pi m_1^2} \left( 1+  O(k^2/m_1^2) \right).
    \label{eq:GG}
\end{align}
Thus, we obtain
\begin{align}
    \Gamma_{\text{vector}}^{(3)} \approx -\frac{1}{8\pi m_1^2} \int_\Sigma d^2x H^I_{ab} H^{Iab}.
\end{align}

Since the fourth term in Eq.~\eqref{eq:expansion-of-K-hat} arises from the mixing between $\tilde{\varphi}^I$ and $a^J_a$ in Eq.~\eqref{eq:I_tilde-varphi}, $\Gamma_{\text{vector}}^{(4)}$ corresponds to the following diagram:
\begin{align}
    \Gamma_{\text{vector}}^{(4)} =
    \begin{tikzpicture}[baseline={(current bounding box.center)}]
    \begin{feynhand}
        \vertex [] (a) at (-1.5,0) {};
        \vertex [] (b) at (1.5,0) {};
        \vertex [ringdot] (v1) at (-0.5,0) {};
        \vertex [ringdot] (v2) at (0.5,0) {};
        \propag [glu] (a) to (v1);
        \propag [glu] (v2) to (b);
        \draw [boson] (v1) arc (180:0:0.5) node[midway, above] {$a_a^I$};
        \draw [boson] (v2) arc (0:-180:0.5) node[midway, below] {$\tilde{\varphi}^J$};
    \end{feynhand}
    \end{tikzpicture}
    &=-\frac{1}{2m_1^2} \int \frac{d^2 k}{(2\pi)^2} H^I_{ab}(k) H^{Iab}(-k) \int \frac{d^2 p}{(2\pi)^2} \frac{p^2}{p^2 + m_1^2} \frac{1}{(p+k)^2 + m_1^2}.
\end{align}
Using $p^2=(p^2+m_1^2)-m_1^2$, the loop integral can be decomposed into
a tadpole term, whose renormalized value is given in Eq.~\eqref{eq:I_ren},
and the finite two-propagator integral in Eq.~\eqref{eq:GG}. 
Keeping the leading term in the low-momentum expansion gives
\begin{align}
    \Gamma_{\text{vector}}^{(4)} \approx \frac{1}{8\pi m_1^2} \left[ 1 - \log \frac{\mu^2}{m_1^2}  \right] \int_\Sigma d^2x H_{ab}^I H^{Iab}.
\end{align}
Setting $\mu=m_1$, we obtain
\begin{align}
    \Gamma_{\text{vector}} \ni \frac{1}{24\pi m_1^2} \int_\Sigma d^2x H_{ab}^I H^{Iab} = \frac{r_0^2}{48\pi} \int_\Sigma d^2x H_{ab}^I H^{Iab}.
\end{align}
Therefore,
\begin{align}
    c_{\mathrm{SO}(3)}^{\text{1-loop}} =  \frac{1}{24\pi m_1^2},
\end{align}
which reproduces Eq.~\eqref{eq:1-loop-level-correction-from-l=1}.

\newcommand{\arxivfont}{\rmfamily}
\bibliographystyle{yautphysm}
\bibliography{ref}

\end{document}